%% file: ws-jai.tex
\documentclass{ws-jai}
\usepackage{xcolor}
\usepackage[flushleft]{threeparttable}
\usepackage{subcaption}
\begin{document}

\catchline{}{}{}{}{} % Publisher's Area please ignore

\markboth{D. Gallardo}{An Ultra-Wideband Dual Polarization Antenna Array for the Detection and Localization of Bright Fast Radio Transients in the Milky Way}

\title{An Ultra-Wideband Dual Polarization Antenna Array for the Detection and Localization of Bright Fast Radio Transients in the Milky Way}

\author{D. Gallardo$^{1, 5}$, R. Finger$^{1, 2}$, F. Solis$^{1}$, D. Monasterio$^{3}$, S. Jorquera$^{2}$, J. Pizarro$^{1}$, J. Riquelme$^{1}$, F. Curotto$^{1}$, F. Pizarro$^{4}$ and L. Bronfman$^{1}$}

\address{
% \\
$^{1}$Department of Astronomy, Universidad de Chile, Santiago, Chile\\
$^{2}$Department of Electrical Engineering, Universidad de Chile, Santiago, Chile\\
$^{3}$National Radio Astronomy Observatory, Charlottesville, Virginia, USA\\
$^{4}$Escuela de Ingeniería Eléctrica, Pontificia Universidad Cat\'olica de Valpara\'iso, Valpara\'iso, Chile\\
$^{5}$\textup{diego.gallardo@raig.uchile.cl}
}

\maketitle

% \corres{$^{2}$Corresponding author.}

\begin{history}
\received{(to be inserted by publisher)};
\revised{(to be inserted by publisher)};
\accepted{(to be inserted by publisher)};
\end{history}

\begin{abstract}

Fast radio bursts (FRBs) are extraordinary astrophysical phenomena characterized by short radio pulses that last only a few milliseconds, yet their power can surpass that of 500 million suns. To date, most detected FRBs originate from beyond our galaxy. However, if an FRB were to originate within the Milky Way, it could be detected using small antennas. In this paper, we propose a compact and \textit{ad-hoc} antenna array designed for the efficient detection and localization of FRBs within the Milky Way. The antenna operates within the 1200-1800~MHz range and consists of three sub-arrays placed in an L-shape for source localization, occupying a total volume of $80 \times 25 \times 6 \textup{ cm}^3$. Each sub-array consists of 4 miniaturized, dual-polarized, half-space radiation antenna elements, forming a one-dimensional array that allows shaping the radiation pattern to match the form of the Milky Way without exhibiting grating lobes. A prototype was constructed and characterized to validate the design. The measured results exhibit good agreement with the simulations. In addition to having a custom elongated radiation pattern, the array has attractive merits, such as low reflections at the input ports, high radiation efficiency, and a distribution that inhibits the existence of phase ambiguities, thus facilitating source localization.

\end{abstract}


\input{intro.tex}
\input{predesign.tex}
\input{structure.tex}
\input{design.tex}
\input{meas.tex}
\input{conclus.tex}

\end{document}

%% file: intro.tex
\section{Introduction}
\label{s: intro}

\noindent In 2007, Lorimer et al. discovered the first fast radio burst (FRB)~\cite{lorimer}, opening up a new realm of exploration for the scientific community. While over 600 FRBs have been published (and over thousands have been detected), their origins and mechanisms remain largely unknown. However, most proposed theories attribute the phenomenon to magnetars~\cite{FRBs 2020}. According to current observations, FRBs manifest as radio waves within the 100--8000~MHz frequency range, exhibiting a brief duration that spans only a few milliseconds. Despite their fleeting nature, these bursts exhibit an extremely bright brightness temperature, $T_\mathrm{B}\gtrsim10^{36}~K$ ~\cite{FRBs 2019}.

Most detected FRBs have been of extragalactic origin, with only one confirmed to be of Galactic origin~\cite{galactic frb}. Owing to the very large distances between their source and Earth, the power received from these events is minute. Consequently, the detection of FRBs has required the utilization of large radio telescopes such as the Canadian Hydrogen Intensity Mapping Experiment (CHIME), Parkes, and the Five-hundred-meter Aperture Spherical Telescope (FAST). However, if an FRB were to originate within the Milky Way, it is estimated that it could be detected using smaller radio telescopes. Motivated by this potential, the STARE-2 experiment (Survey for Transient Astronomical Radio Emission 2) was conceived~\cite{stare2}, leveraging a modest antenna design. For over 1000 days, STARE-2 has been diligently scanning for FRBs within our galaxy. It detected many solar bursts and one FRB associated with the known Galactic magnetar SGR~$1935+2154$ (which was also detected by the CHIME/FRB collaboration~\cite{chime galactic}).

The ARTE project (Astronomical Radio Transients Experiment) is a low-cost radio telescope under development in the Millimeter Wave Laboratory of the Universidad de Chile, whose objective is to detect and give an approximate location of FRBs within the Milky Way. What sets ARTE apart is its cost-effectiveness and its goal to be a more sensitive version of STARE-2 while operating from a single site. High precision in the location of the FRBs is not expected, but a resolution similar to the 15 deg$^2$ of STARE-2 is sought. A key factor contributing to ARTE's affordability is the simplicity of its components, particularly the antenna, which costs less than 250~USD in manufacturing materials, with a total volume of under $80 \times 25 \times 6 \textup{ cm}^3$.

\begin{figure}[t!]
    \centering
    \includegraphics[width = 0.4\linewidth]{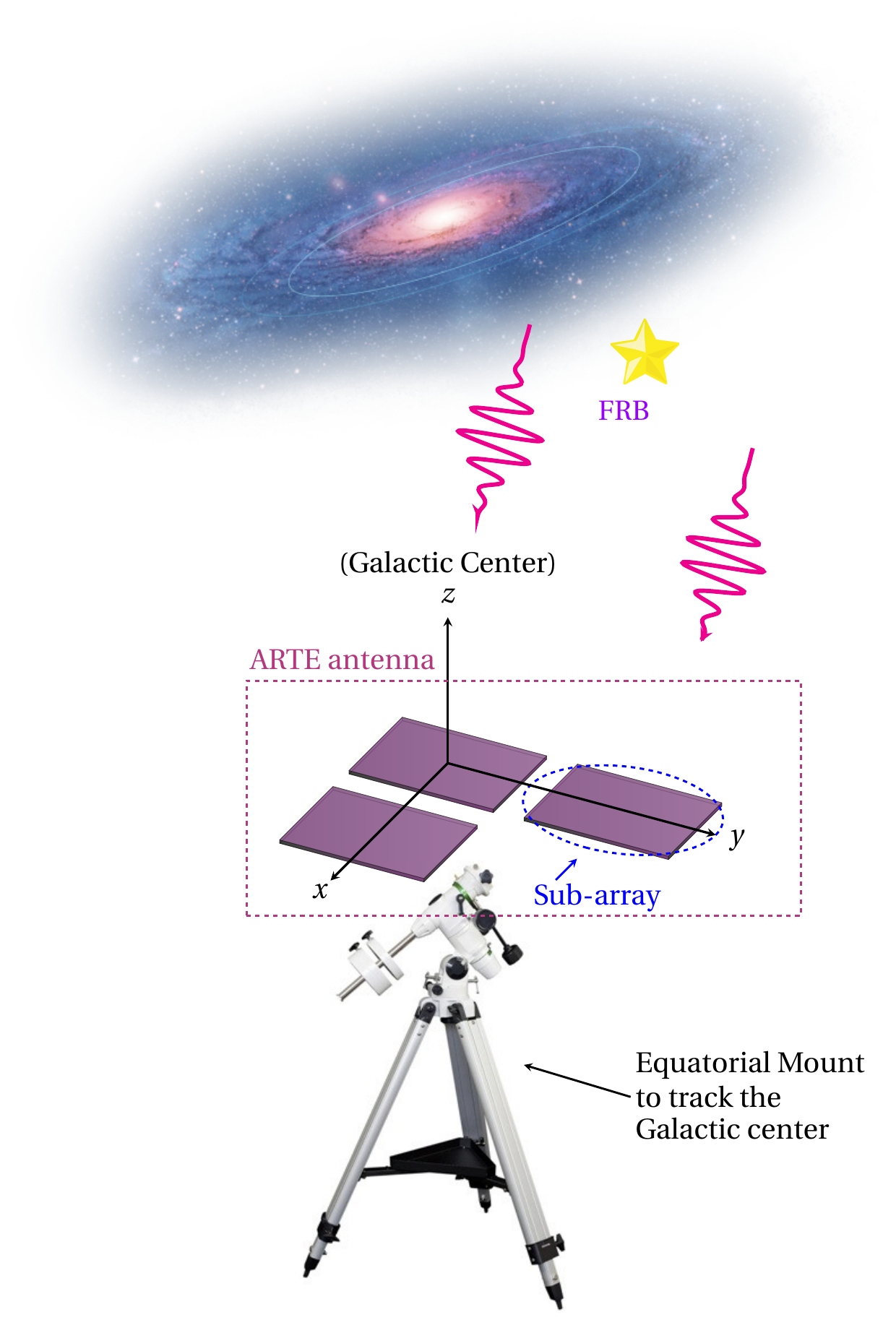}
    \caption{Global scheme of the proposed antenna array for the ARTE project.}
    \label{fig: initial concept}
\end{figure}

This paper presents an innovative, cost-effective, compact, and \textit{ad-hoc} antenna array designed for efficient detection and localization of FRBs within the Milky Way. Operating between 1200--1800~MHz, the ARTE antenna array employs an equatorial mount so that the beam pattern keeps aligned with the Galactic plane, while the Galactic center is tracked, as depicted in Fig.~\ref{fig: initial concept}. The frequency band of 1200–-1800~MHz has been selected for a few reasons: compactness of the overall array (at these frequencies), reduced manufacture cost and ease of implementation, largely known detection of FRBs in the extragalactic frequency spectrum, and evidence of earlier detections at these frequencies from an FRB-like burst~\cite{stare2}. Structurally, the array integrates three sub-arrays arranged in an L-shaped configuration. This arrangement enables source localization in the $x$ and $y$ axes, aligning with antenna theory principles~\cite{antenna theory}. Each of the sub-arrays comprises a set of four dual-polarized antenna elements. The rationale behind this arrangement is detailed in Section~\ref{s: design considerations}, but in the first instance, we can say that it allows us to achieve an elongated radiation pattern aligned with the Galactic plane, thereby increasing sensitivity on our primary area of interest.

%% file: predesign.tex
\section{Design Considerations}
\label{s: design considerations}

The specifications for the ARTE antenna are outlined in Table~\ref{table: specs} and are discussed in detail in this section.

\begin{table}[t!]
    \caption{ARTE antenna specifications.}
    \centering
    \scalebox{0.95}[0.95]{%
    \begin{tabular}{l|l|c}
    \hline \hline
    & \textbf{Parameter} & \textbf{Value} \\ \hline
    1) & Frequency band & 1200--1800~MHz \\ 
    2) & $S_{11}$ & $<-10$~dB \\
    3) & Simulated Radiation Efficiency & $>90\%$ \\ 
    4) & Dual-pol? & Yes \\
    5) & HPBW$_{x}$ & $>80^\circ$ \\
    6) & HPBW$_{y}$ & $<15^\circ$ \\
    7) & Front-to-back ratio & $>10$~dB \\
    8) & Ability to localize sources? & Yes \\ \hline \hline 
    \end{tabular}
    }
    \label{table: specs}
\end{table} 

\subsection{Reflection coefficient and Radiation Efficiency}

To efficiently couple the FRB signal with the rest of the instrument while minimizing added noise, the antenna's reflection coefficient, represented by the S-parameter $S_{11}$, must be low, and the radiation efficiency must approach 100\%. Since the antenna must satisfy other more stringent criteria, we have chosen the standard thresholds outlined in Table~\ref{table: specs} for the $S_{11}$ parameter and the radiation efficiency. 

\subsection{Dual Polarization}

As documented by Petroff in~\cite{frb}, FRBs may exhibit a certain degree of polarization. Given the uncertainty regarding the source's polarization type or specific orientation, employing a dual-polarization (or \textit{dual-pol}) antenna becomes necessary. This implies that the three sub-arrays must have two outputs, each corresponding to an orthogonal polarization.

\subsection{Radiation Pattern and Half-Power Beamwidth (HPBW)}

\begin{figure}[t!]
    \centering
    \begin{subfigure}[b]{0.5\linewidth}
         \centering
         \includegraphics[width = 0.75\linewidth]{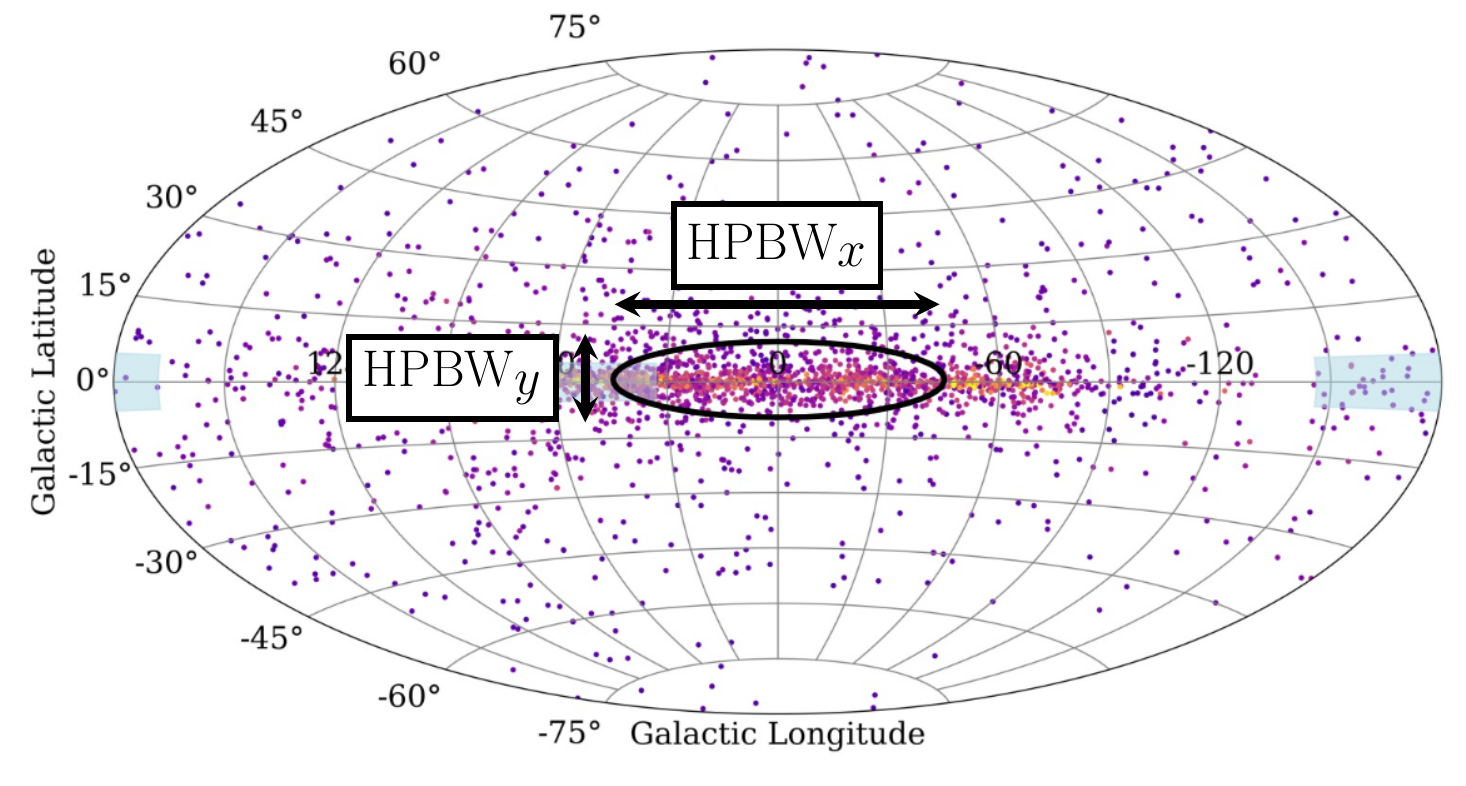}
         \caption{}
     \end{subfigure}
     \begin{subfigure}[b]{0.19\linewidth}
         \centering
         \includegraphics[width = 0.75\linewidth]{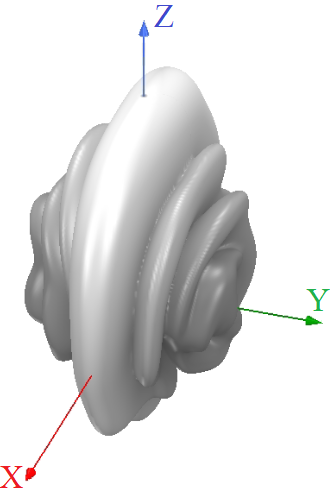}
         \caption{}
     \end{subfigure}
     \begin{subfigure}[b]{0.29\linewidth}
         \centering
         \includegraphics[width = 0.75\linewidth, trim = {14cm 0 13cm 0}, clip]{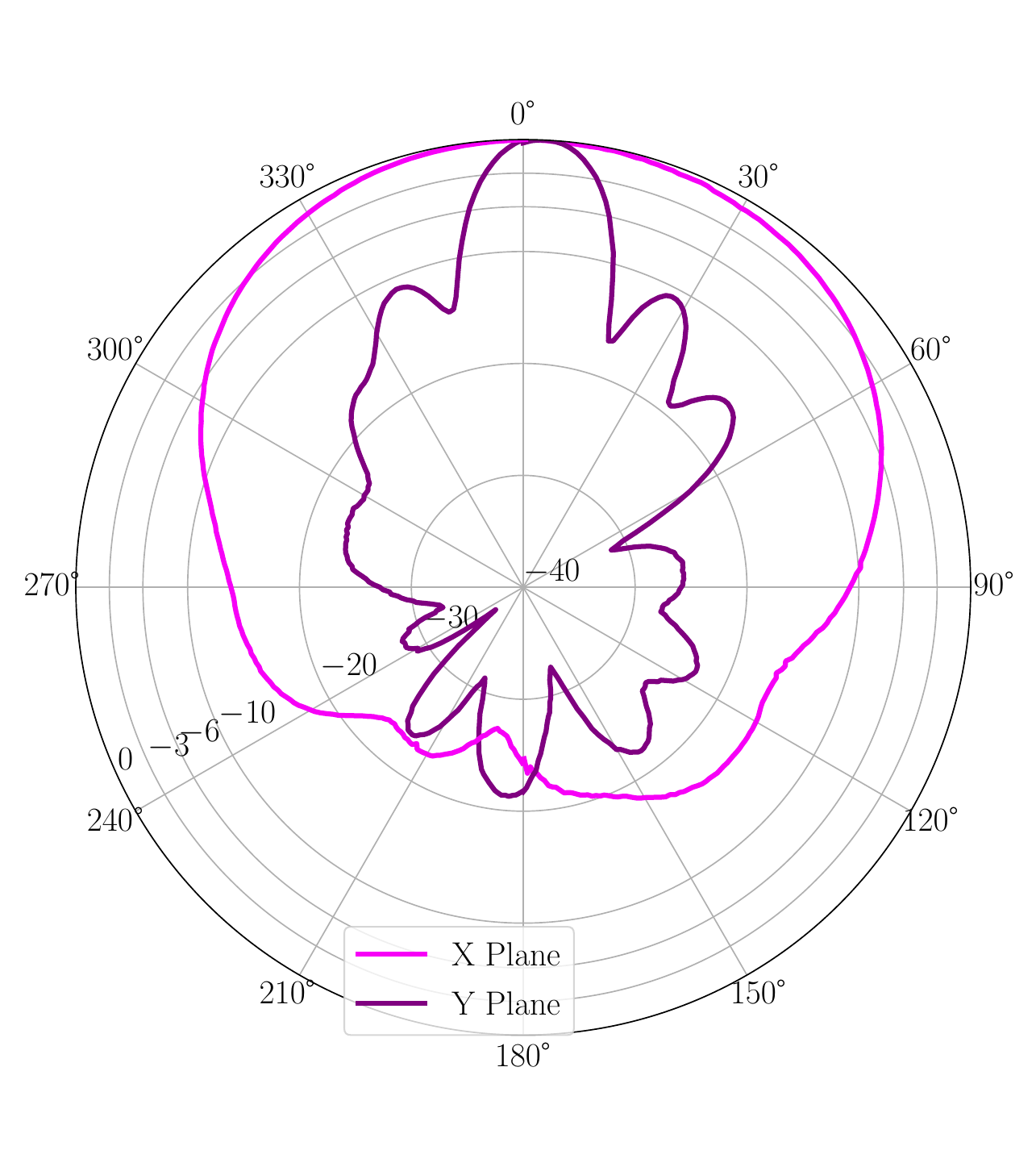}
         \caption{}
     \end{subfigure}
    \caption{Expected shape of the radiation pattern of the ARTE antenna. (a) Projection in the Galactic plane~\cite{pulsar distribution}. (b) 3D radiation pattern. (c) Cuts in the principal planes. All radiation patterns are plotted at 1.5~GHz.}
    \label{fig: pat shape}
\end{figure}

\begin{figure}[t!]
    \centering
     \begin{subfigure}[h!]{0.49\linewidth}
         \centering
         \includegraphics[width = 1\linewidth]{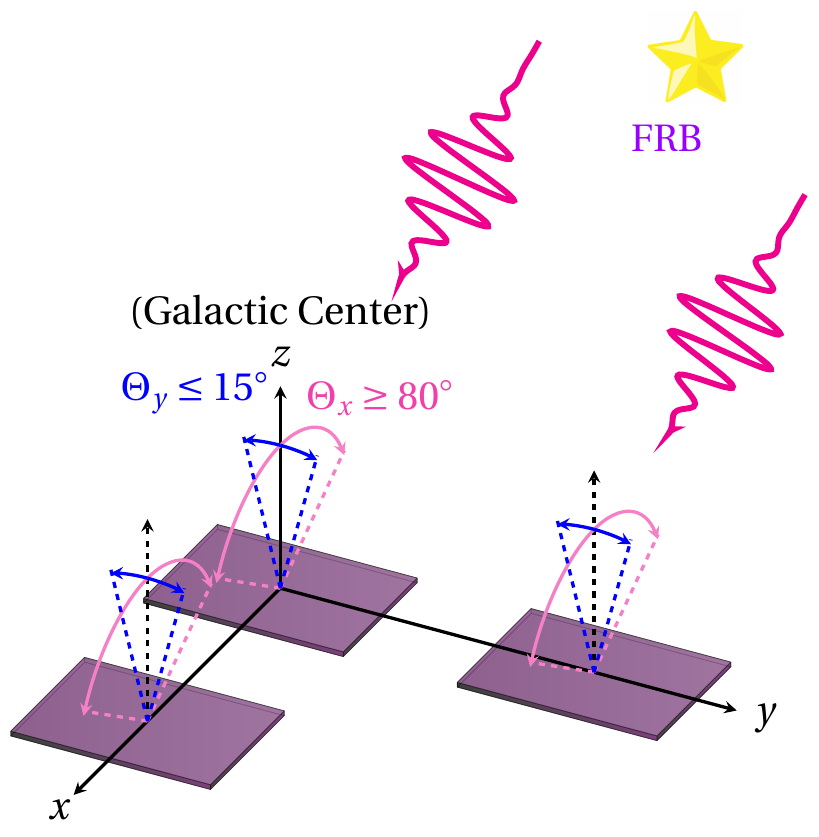}
         \caption{}
     \end{subfigure}
     \begin{subfigure}[h!]{0.49\linewidth}
         \centering
         \includegraphics[width = 1\linewidth]{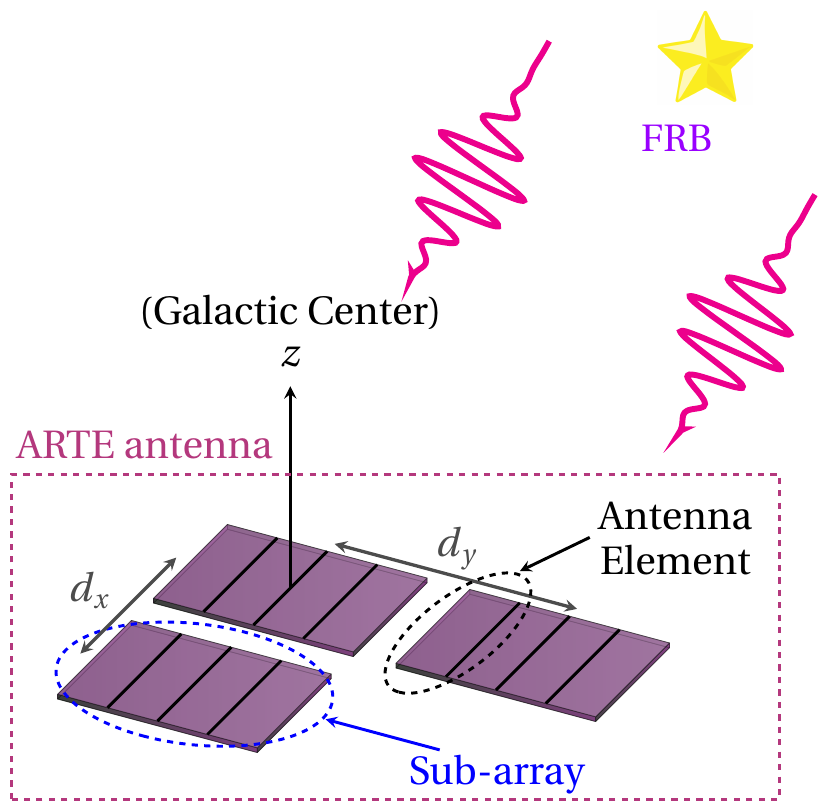}
         \caption{}
     \end{subfigure}
    \caption{Specifications of the ARTE antenna. (a) Each sub-array is required to have an HPBW of at least $80^\circ$ along the $x$ axis ($\Theta_x$) and at most $15^\circ$ along the $y$ axis ($\Theta_y$). (b) Further subdivision of each sub-array into four antenna elements for radiation pattern shaping.}
    \label{fig: generic array}
\end{figure}

As discussed in Section~\ref{s: intro}, the primary objective of the ARTE project is to observe FRBs originating from the Milky Way. Consequently, the ARTE antenna is designed to observe the Milky Way exclusively. This allows us to form an initial understanding of the expected radiation pattern. When examining the projection of the radiation pattern in the Galactic plane, the anticipated result resembles what is shown in Fig.~\ref{fig: pat shape}(a). Furthermore, considering the complete pattern in three dimensions, the expected representation is illustrated in Fig.~\ref{fig: pat shape}(b). At the same time, the cuts in the principal planes are shown in Fig.~\ref{fig: pat shape}(c). The points shown in Fig.~\ref{fig: pat shape}(a) correspond to the distribution of pulsars in the Galactic plane, while the ellipse corresponds to the projection of the radiation pattern. The axes of the ellipse correspond to the HPBWs on the $x$ and $y$ axes. Given the distribution of pulsars and expecting a similar distribution for FRBs (just as an ansatz), we can define the limits for the HPBWs as HPBW$_x > 80^\circ$ and HPBW$_y < 15^\circ$, as shown in Fig.~\ref{fig: generic array}(a). These thresholds ensure that most of the antenna gain is concentrated in the sky region of our primary interest.

We implemented a one-dimensional antenna array to achieve the desired radiation pattern, as shown in Fig.~\ref{fig: generic array}(b).  The selection of the antenna element and its incorporation into a four-antenna array are elaborated upon in Section~\ref{s: structure}, while a comprehensive design breakdown is provided in Section~\ref{s: design}. 

\subsection{Direction of Arrival (DoA) Estimation: L-shaped Array and Digital Combination}  
\label{ss: doa}

\begin{figure}[t!]
    \centering
     \begin{subfigure}[B]{0.65\linewidth}
         \includegraphics[width = 1\linewidth]{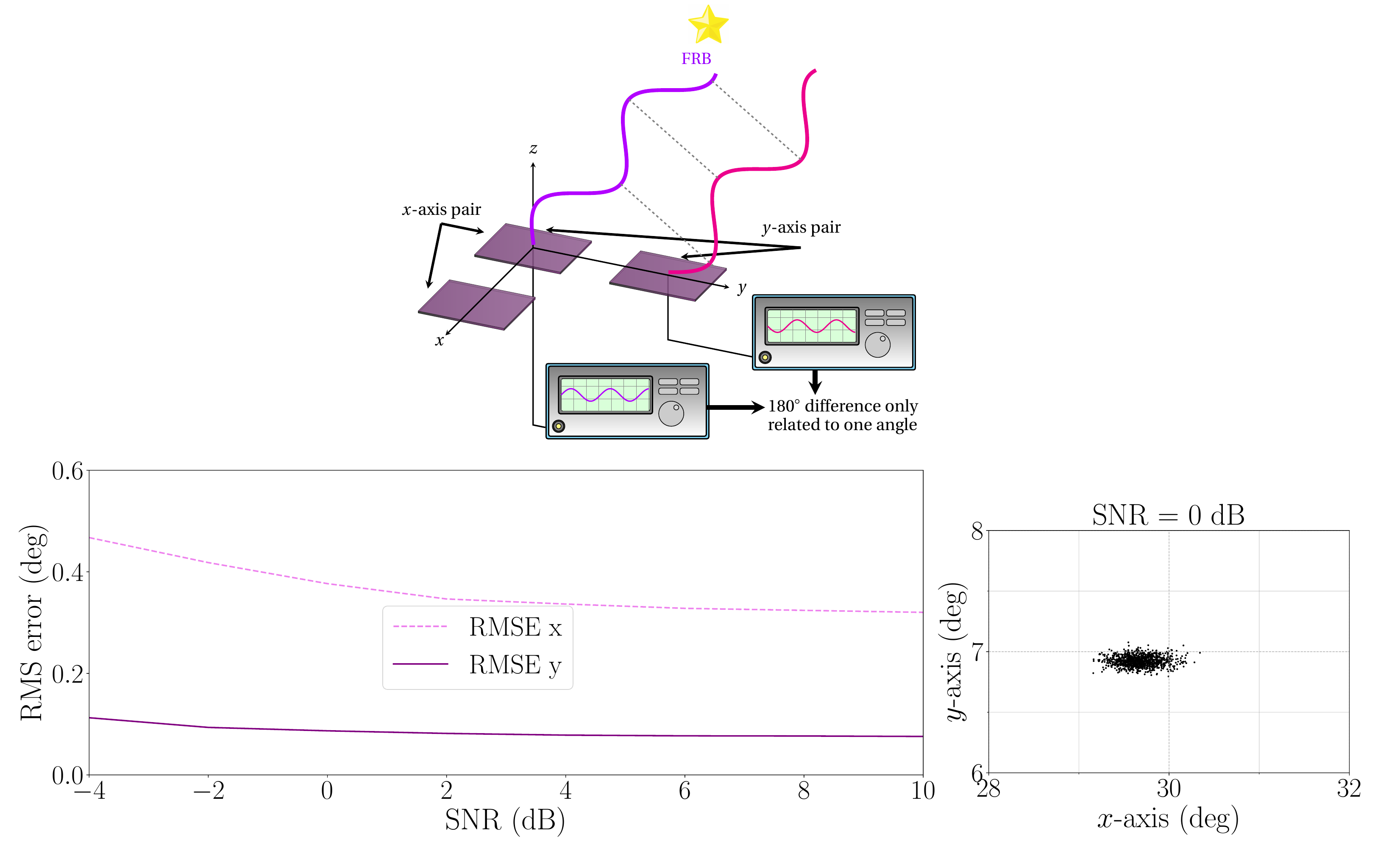}
         \caption{}
     \end{subfigure}
     \begin{subfigure}[B]{0.34\linewidth}
         \includegraphics[width = 1\linewidth]{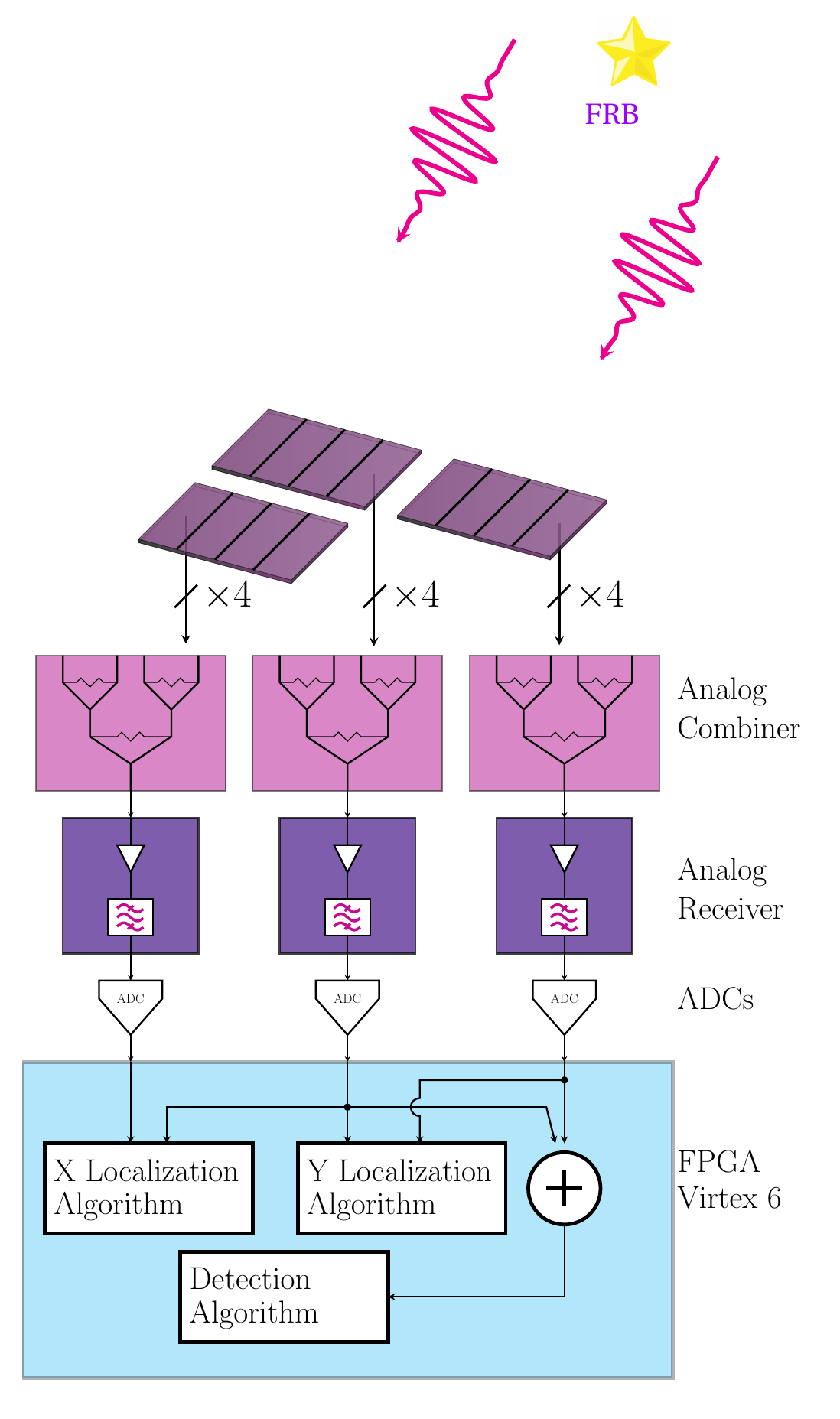}
         \caption{}
     \end{subfigure}
    \caption{Representation of detection and localization within the ARTE experiment. (a) Signal localization is done by measuring phase shifts using the U-ESPRIT 2D algorithm. Simulations were conducted considering an accumulation time of $50~\mu\textup{s}$, an FFT of 2048 channels, and an ADC sampling rate of 1.2~GSPS. The RMSE of the estimated angles and the scatter plot for a source with SNR $= 0$~dB located at $(30^\circ, 7^\circ)$ are illustrated. A total of 1000 trial runs were considered. (b) Simplified representation of the analog receiver and the digital architecture.}
    \label{fig: localization and detection}
\end{figure}

The three sub-arrays are configured in an L-shape arrangement to facilitate source localization. This configuration results in two antenna pairs: one along the $x$-axis and another along the $y$-axis, with a shared antenna. By measuring the phase shift of an incoming signal across these paired sub-arrays, the direction of arrival can be determined\footnote{Assuming that the array is constructed without phase ambiguities.}~\cite{antenna theory}. This concept is illustrated with an example in Fig.~\ref{fig: localization and detection}(a), where the phase difference is only related to one angle of arrival (that is, there is a bijective function between phase difference and angle of arrival). For ARTE, the plan is to utilize sub-space methods for measuring the phase difference, such as MUSIC~\cite{music}, ESPRIT~\cite{esprit}, or U-ESPRIT 2D~\cite{uesprit2d}. These methods, known for their reliability and consistent performance over the years, have achieved source localization accuracy of less than one degree, even with signal-to-noise ratios less than 0~dB and antenna array sizes as small as $4\lambda$~\cite{uesprit2d, doaexp1, doaexp2}, which results in better performance than the $\lambda / D$ angular resolution expected from traditional beam synthesis. 
For the specific case of ARTE, we conducted simulations to estimate the expected error in source localization. Specifically, we simulated the 2D U-ESPRIT algorithm, considering the incidence of a quasi-monochromatic source with amplitude and phase modulation. We configured the array in the L-shape depicted in Fig.~\ref{fig: generic array}(b), with the distances $d_x =$~8.3~cm and $d_y =$~32.2~cm, discussed in Section~\ref{ss: separation}. The root-mean-square error (RMSE) as a function of SNR and a scatter plot of the estimated locations are shown in Fig.~\ref{fig: localization and detection}(a). We anticipate an angular error of approximately one deg$^2$ from the simulations. However, it is essential to note that these simulations do not account for real-life effects, such as phase and amplitude errors resulting from imperfections in the antennas and receiving system and multipath effects caused by ground reflections and scattering.

The preceding paragraph discussed signal localization through phase difference measurement, a conventional technique. Yet, to detect signals —especially FRBs— an analog receiver equipped with an amplification chain is imperative. Moreover, the signals must undergo digitization via Analog-to-Digital Converters (ADCs), as illustrated in Fig.~\ref{fig: localization and detection}(b). Once we digitize the incoming signal, we can replicate the signals from the three sub-arrays without compromising the signal-to-noise ratio. We can then allocate one set of these signals for localization algorithms such as U-ESPRIT 2D while utilizing the other to combine the outputs of the two $y$-axis sub-arrays for the detection algorithm. This approach effectively forms an equivalent 8-antenna array in the $y$-axis for detecting FRBs, enhancing overall gain and sensitivity in our area of interest while maintaining the antenna's size and complexity.

\subsection{The Size of the Antenna}
\label{ss: separation}

\begin{figure}[t!]
    \centering
     \begin{subfigure}[h!]{0.24\linewidth}
         \includegraphics[width = 1\linewidth]{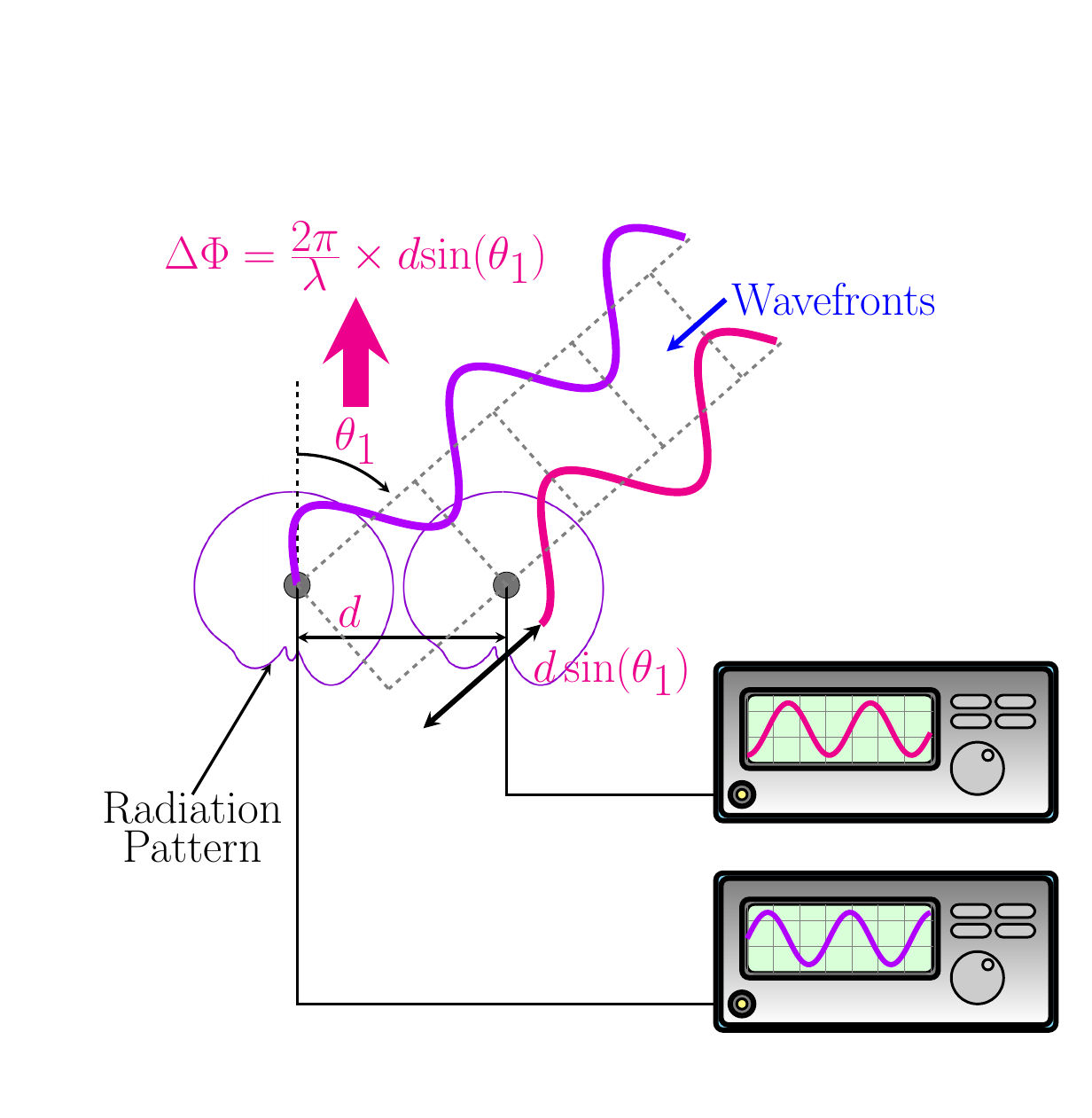}
         \caption{}
     \end{subfigure}
     \begin{subfigure}[h!]{0.24\linewidth}
         \includegraphics[width = 1\linewidth]{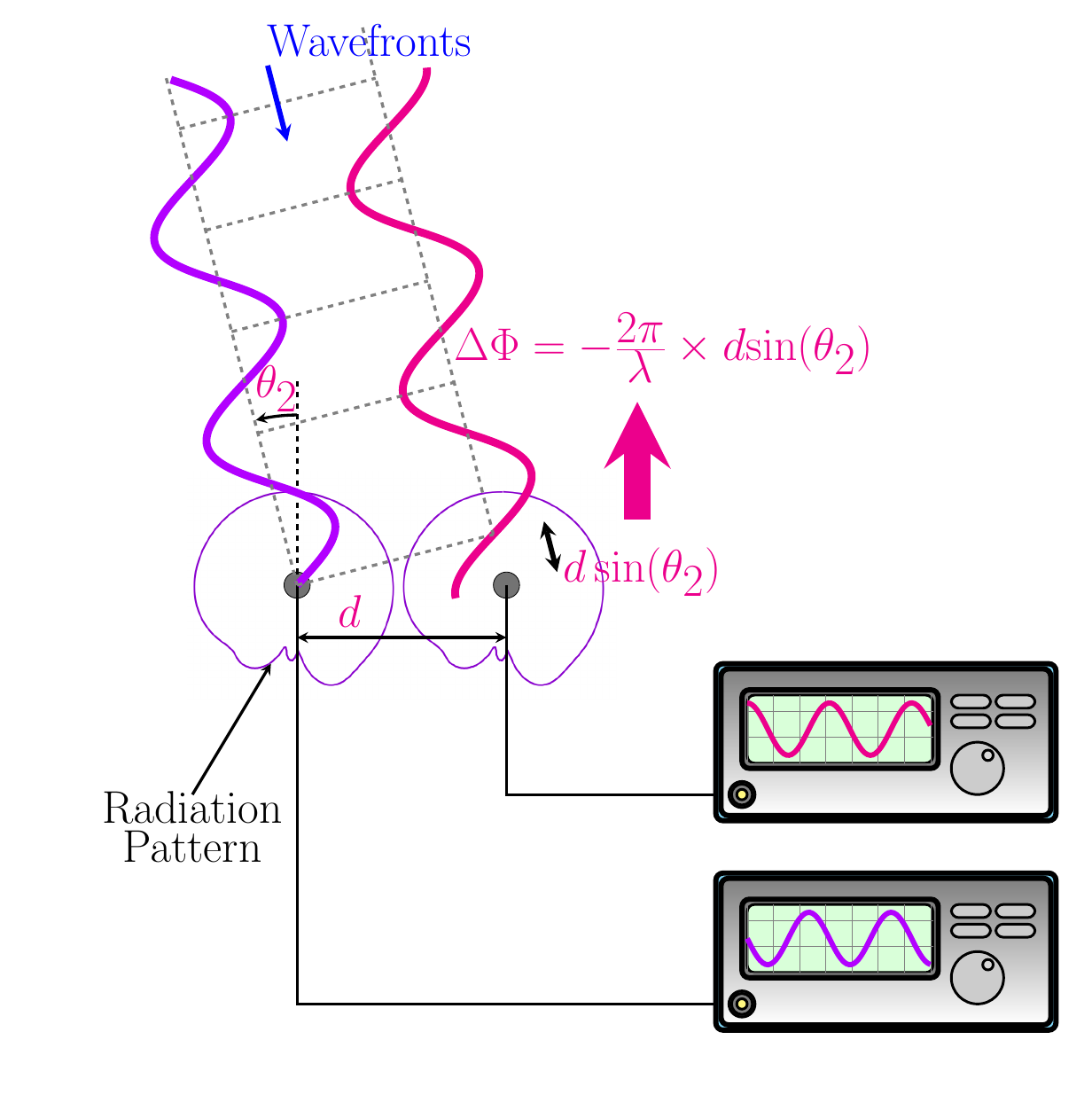}
         \caption{}
     \end{subfigure}
     \begin{subfigure}[h!]{0.24\linewidth}
         \includegraphics[width = 1\linewidth]{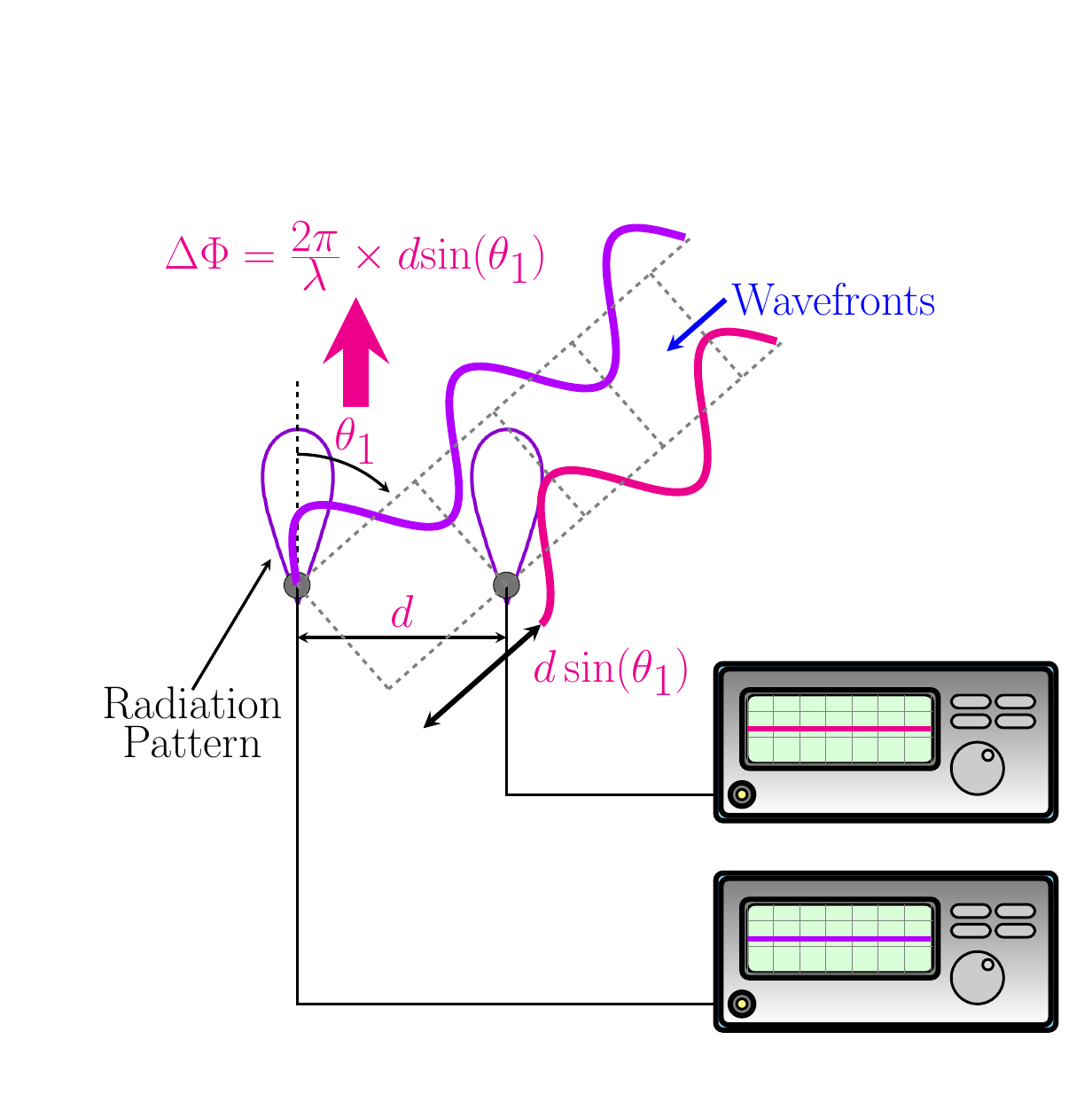}
         \caption{}
     \end{subfigure}
     \begin{subfigure}[h!]{0.24\linewidth}
         \includegraphics[width = 1\linewidth]{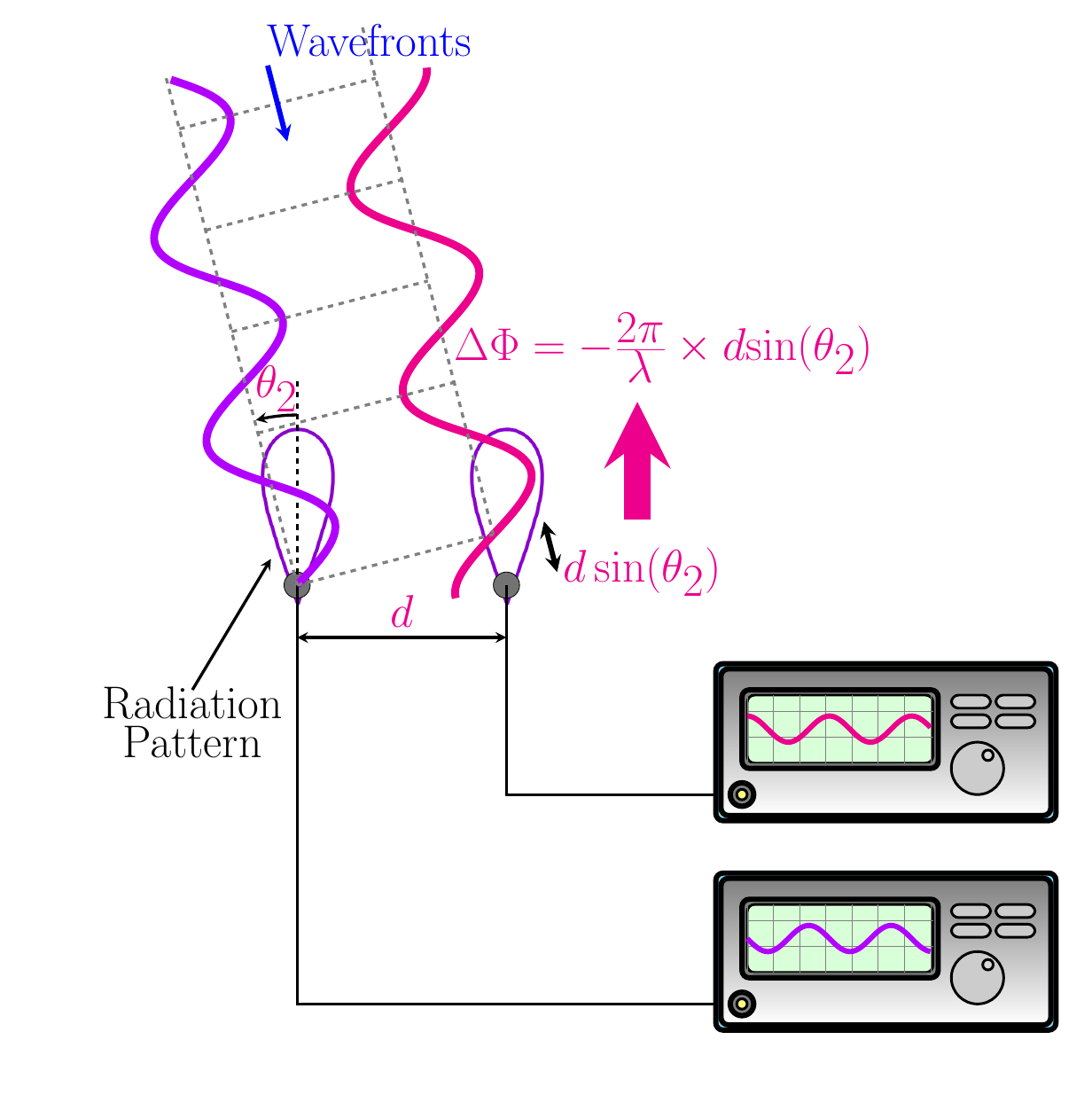}
         \caption{}
     \end{subfigure}
    \caption{Graphical representation of phase ambiguity. In (a) and (b), where the radiation pattern is wide and $d > \lambda / 2$, the phase difference between antennas remains the same for angles $\theta_1$ and $\theta_2$. Conversely, in (c) and (d), where the radiation pattern is narrow, the antennas are not sensitive to angle $\theta_1$. Consequently, there is no phase ambiguity between angles $\theta_1$ and $\theta_2$.}
    \label{fig: phase ambig}
\end{figure}

It is well known that to form a uniform and linear array without phase ambiguities, the antennas comprising the array should be spaced by $\lambda / 2 = c / (2f)$, being $c$ the speed of light, $f$ the operation frequency and $\lambda$ the wavelength~\cite{antenna theory}. The reason behind this criteria is that, referring to Figs.~\ref{fig: phase ambig}(a) and (b), if we want there to be no phase ambiguity in any pair of angles $\theta_1$ and $\theta_2$, we must satisfy 

\begin{align}
    \frac{2\pi d \sin \theta_1}{\lambda} \pm \frac{2\pi d \sin \theta_2}{\lambda} < 2\pi \nonumber \\
    \Leftrightarrow \frac{d}{\lambda}(\sin \theta_1 \pm \sin \theta_2) < 1. \label{eq: no phase amb}
\end{align}
Since~\eqref{eq: no phase amb} must be true for every $\theta_1$, $\theta_2$, and the worst case scenario is when $\sin \theta_1 - \sin \theta_2 = 2$, we conclude that the criteria $d < \lambda/2$ must be satisfied so that there are no phase ambiguities. However, this is only true when the antenna element has a wide radiation pattern. Suppose the antenna element has a narrow radiation pattern, as the one shown in Figs.~\ref{fig: phase ambig}(c) and (d). In that case, there are angles for which phase ambiguity is irrelevant since the antenna is not sensitive at such angles. Assuming that the radiation pattern is null outside the range $\theta \in (-\Theta_0, \Theta_0)$, then $\theta_1$ and $\theta_2$ can only vary within the range $(-\Theta_0, \Theta_0)$, and, therefore, the worst case scenario of~\eqref{eq: no phase amb} changes to $\sin \theta_1 - \sin \theta_2 = 2\sin \Theta_0$. With such a case, the criteria for no phase ambiguities is given by 

\begin{align}
    d < \frac{\lambda}{2\sin \Theta_0}. \label{eq: no phase amb 2}
\end{align}

Since the antenna is expected to have a wide radiation pattern in the XZ plane ($> 80^\circ$) and a narrow one in the YZ plane ($< 15^\circ$), the sub-arrays must be separated by distances $d_x < \lambda / 2$ in the $x$-axis and $d_y < \lambda / (2\sin \Theta_0)$ in the $y$-axis. Also, these criteria must be met for every frequency within the range 1200--1800~MHz. That is to say, the criteria $d_x < \lambda / 2$ and $d_y < \lambda / (2\sin \Theta_0)$ must be met for the smallest $\lambda$, corresponding to 1800~MHz. Finally, if we consider $\Theta_0 = 15^\circ$, we conclude that the antennas must be separated by at most $d_x = $ 8.3~cm and $d_y = $ 32.2~cm. This, in turn, defines the maximum size of the antenna.

\subsection{Front-to-Back Ratio}
\label{ss: fbr}

\begin{figure}[t!]
    \centering
    \includegraphics[width = 0.26\linewidth]{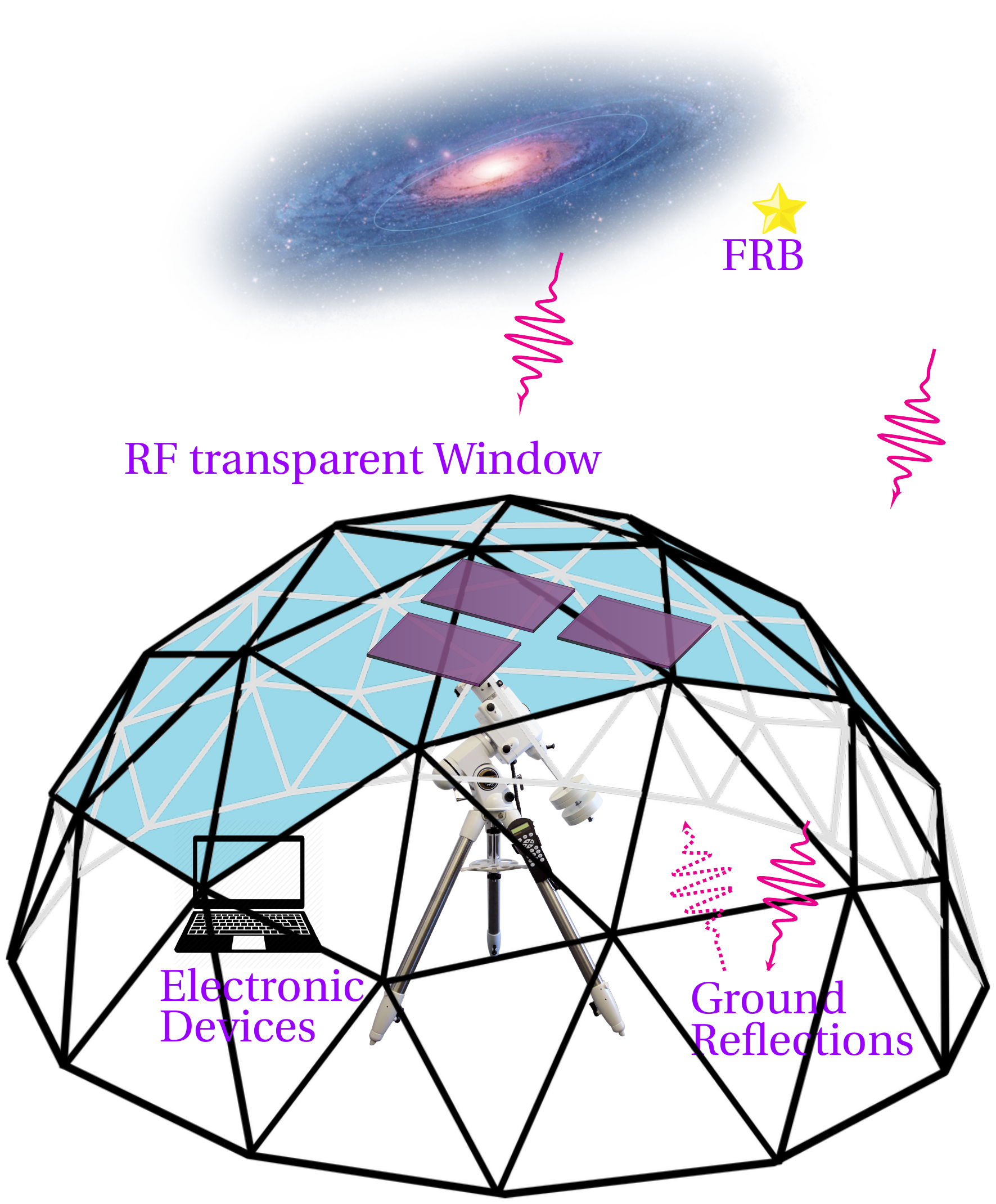}
    \caption{Positioning of the antenna within a dome. The dome has a window that is transparent to RF signals and is incorporated to facilitate observations.}
    \label{fig: dome}
\end{figure}

Since ARTE is intended to operate continuously throughout the year, the antenna array and electronics must be protected within a dome, as shown in Fig.~\ref{fig: dome}. Additionally, the antenna must exhibit a half-space pattern, quantified by the front-to-back ratio (FBR), to increase its gain and, in turn, avoid reflections from the ground and possible interference from nearby electronic equipment. In this study, we have adopted an FBR $> 10$~dB criterion. We can estimate the ground pick based on this FBR criterion. To calculate an exact value, we would need the antenna's 3D radiation pattern, which we do not have. However, we can make estimations using certain assumptions. Firstly, we will assume that the ground is at a constant temperature of 290~K. Secondly, we are going to consider for this estimation that the FBR consists of the ratio between the integrated main lobe and the integrated back lobe, that is,

\begin{align}
    \textup{FBR} = \frac{\int_{\textup{main lobe}} D(\theta, \phi) d\Omega}{\int_{\textup{back lobe}} D(\theta, \phi) d\Omega}, \label{ec: fbr}
\end{align}
being $D(\theta, \phi)$ the directivity pattern in terms of the angular coordinates $\theta$ and $\phi$. Considering all the assumptions above, we obtain a ground pick of~\cite{pozar}

\begin{align}
    T_{\textup{back}} = \frac{\int_{\textup{back lobe}} \overbrace{T(\theta, \phi)}^{290~\textup{K}} D(\theta, \phi) d\Omega}{\int_{\textup{all space}} D(\theta, \phi) d\Omega} = 290~\textup{K}\times \frac{\int_{\textup{back lobe}} D(\theta, \phi) d\Omega}{\int_{\textup{main lobe}} D(\theta, \phi) d\Omega + \int_{\textup{back lobe}} D(\theta, \phi) d\Omega}. \label{ec: temperature 1}
\end{align}
Then, using~\eqref{ec: fbr} with FBR $=$~10, we conclude

\begin{align}
    T_{\textup{back}} = 290~\textup{K}\times\frac{\int_{\textup{back lobe}} D(\theta, \phi) d\Omega}{10\int_{\textup{back lobe}} D(\theta, \phi) d\Omega + \int_{\textup{back lobe}} D(\theta, \phi) d\Omega} \approx 26~\textup{K}. \label{ec: temperature 2}
\end{align}
This ground pick translate into an increase of 26~K in the system's temperature $T_{\textup{sys}}$.

%% file: structure.tex
\section{Antenna Structure}
\label{s: structure}

% Aquí hay que hablar de que la antena se forma por dipolos cruzados, y que conviene ponerlos en diagonal al eje del arreglo para que haya simetría en los patrones de radiación.

As highlighted in Section 2, the decision was made to use an array of 4 antenna elements to make each sub-array. These antenna elements must be dual-pol, and their size should be minimized to avoid phase ambiguities. Given that dual-pol antennas are essentially combinations of two antennas with crossed polarizations, selecting antennas capable of overlapping their centers is advantageous due to the size requirement. Examples of suitable antennas include dipoles or log-periodic antennas~\cite{lpda}. For this work, we used crossed dipoles as the array element~\cite{base station antenna}.

\begin{figure}[t!]
    \centering
    \begin{subfigure}[b]{0.38\linewidth}
         \centering
         \includegraphics[width = 1\linewidth]{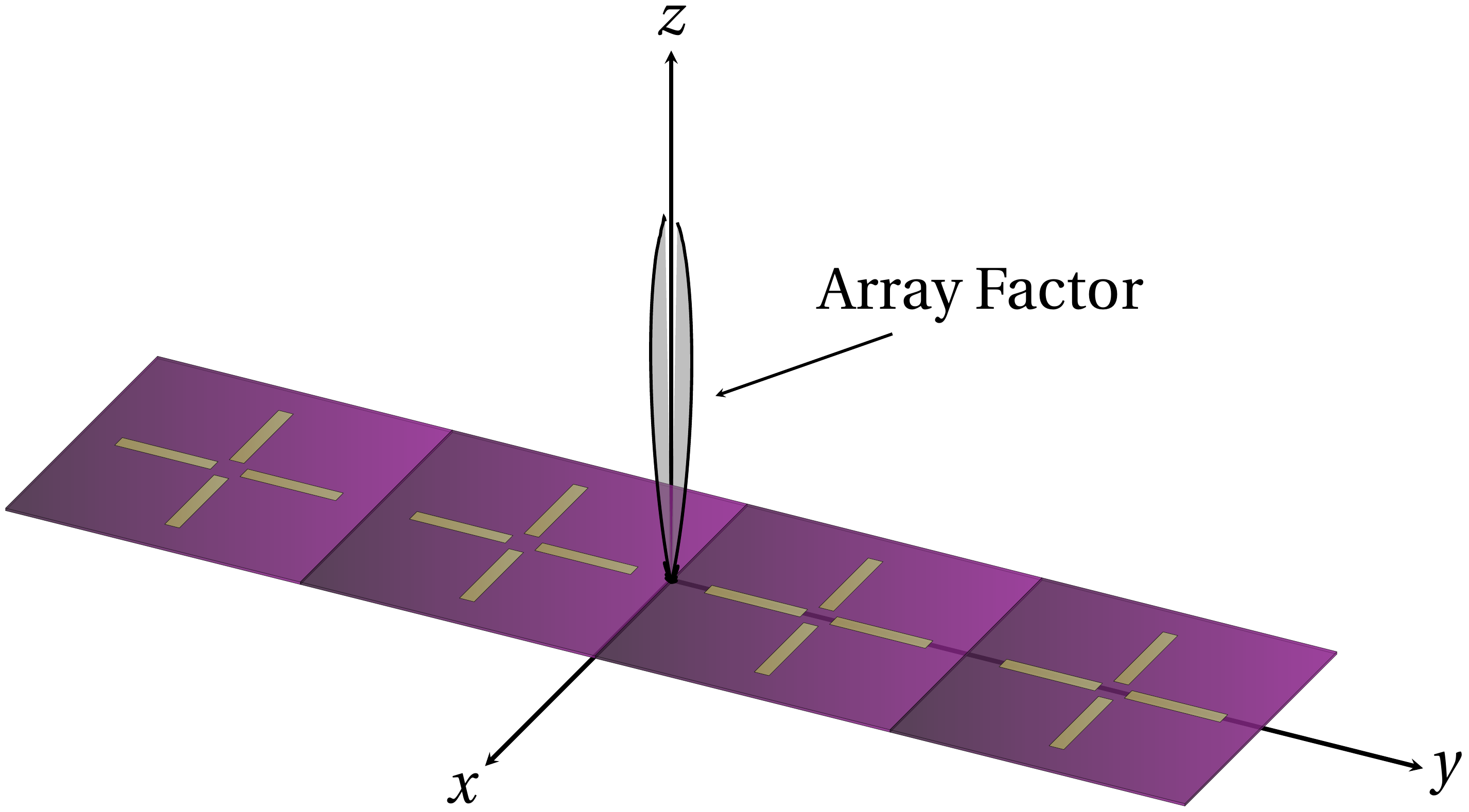}
         \caption{}
     \end{subfigure}
     \begin{subfigure}[b]{0.54\linewidth}
         \centering
         \includegraphics[width = 1\linewidth]{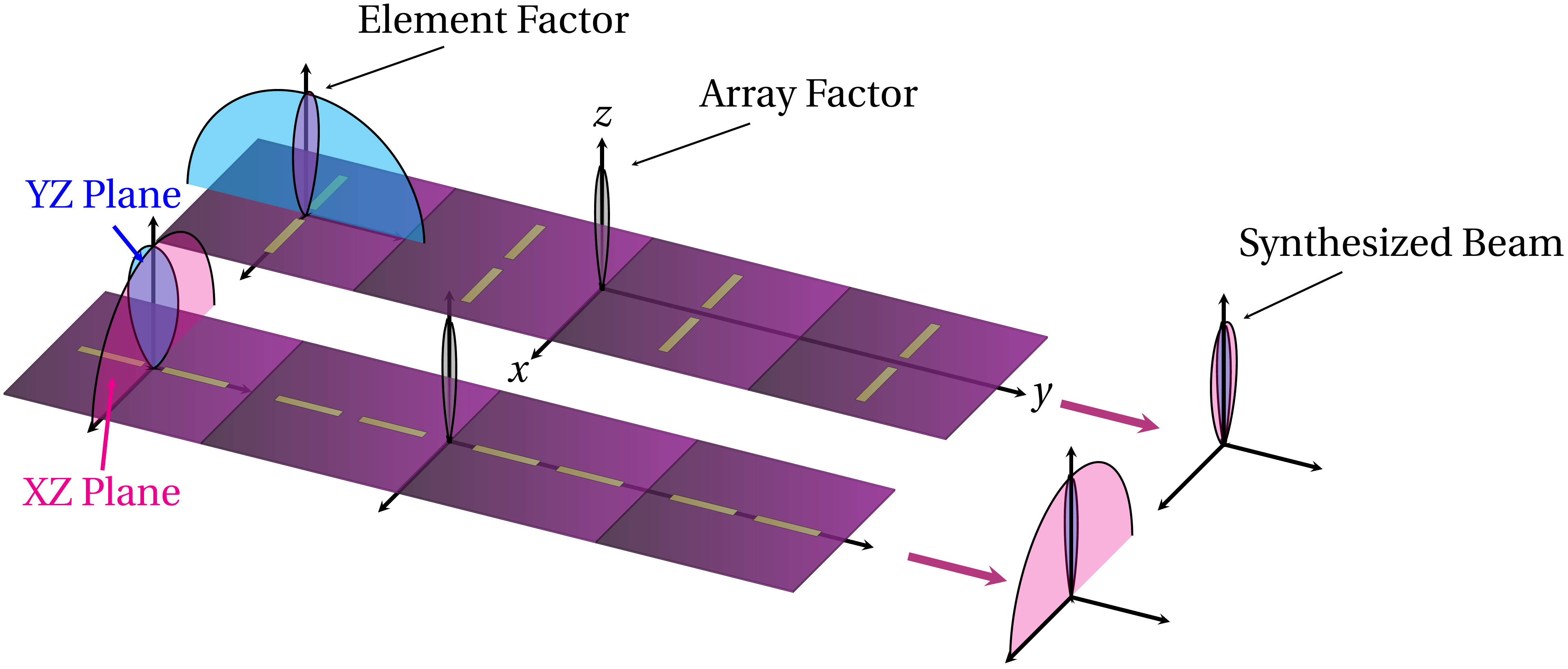}
         \caption{}
     \end{subfigure}
     \begin{subfigure}[b]{0.38\linewidth}
         \centering
         \includegraphics[width = 1\linewidth]{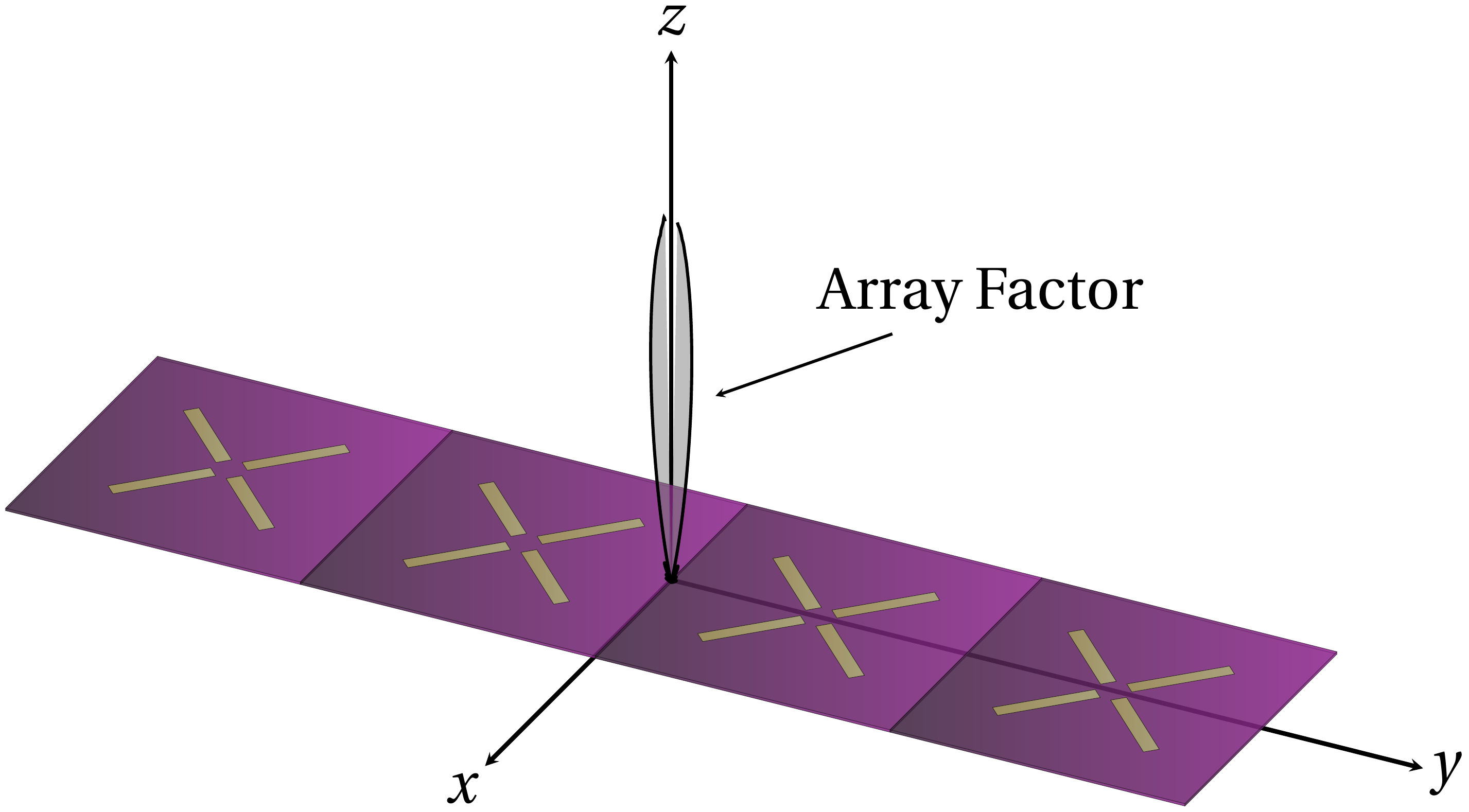}
         \caption{}
     \end{subfigure}
     \begin{subfigure}[b]{0.54\linewidth}
         \centering
         \includegraphics[width = 1\linewidth]{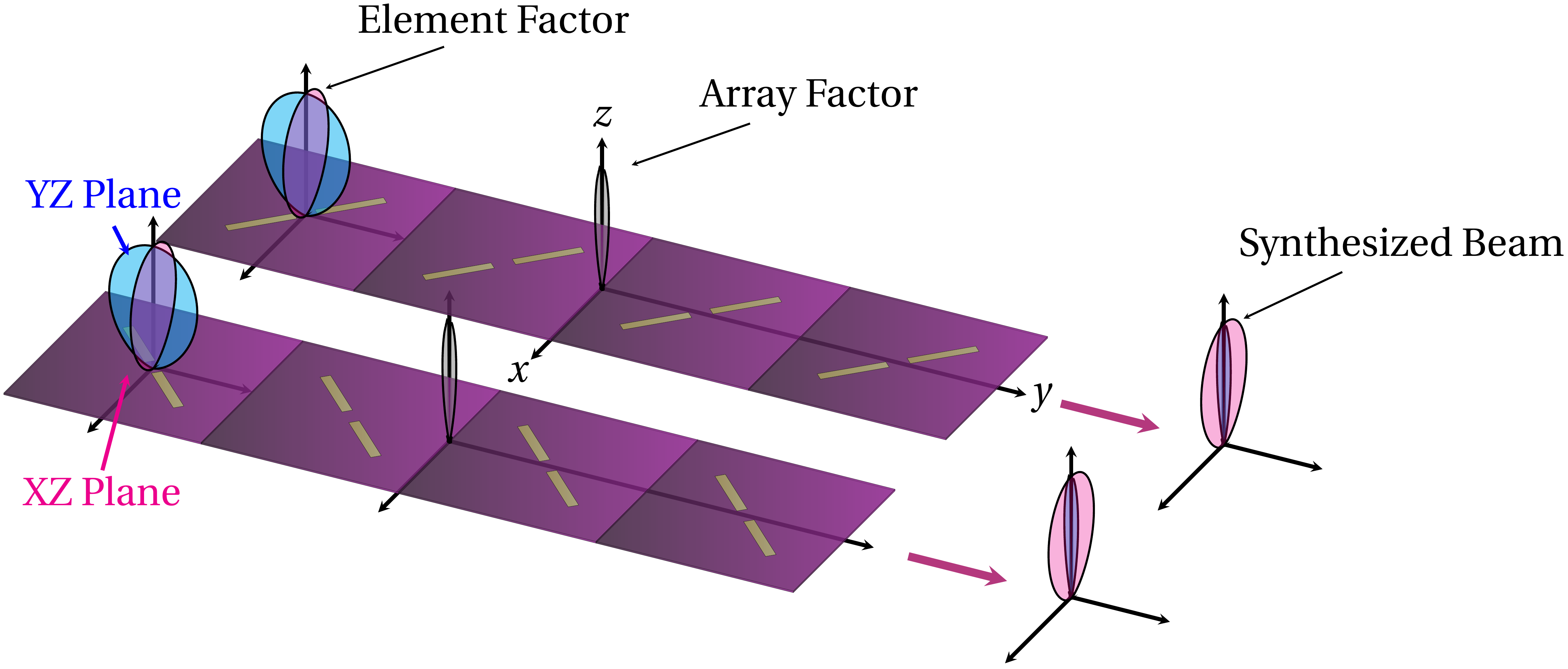}
         \caption{}
     \end{subfigure}
    \caption{Dipole placement in the sub-arrays. (a) and (b) When the dipoles are placed longitudinally, the array factor affects each polarization differently. (c) and (d) When the dipoles are placed diagonally, the array factor affects both polarizations equally.}
    \label{fig: dualpol array}
\end{figure}

However, it's important to highlight a minor limitation when using dipoles in our application. Due to inherent differences in the radiation patterns along the X and Y planes of these dipoles, the resultant antennas exhibit distinct radiation patterns for each polarization once the array is formed, as illustrated in Figs.~\ref{fig: dualpol array}(a) and (b). Such disparity wouldn't arise if the antenna elements maintained consistent patterns across both axes. To solve this problem, we decided to arrange the dipoles diagonally, as depicted in Fig.~\ref{fig: dualpol array}(c). By doing so, we achieve symmetry across the X and Y planes, ensuring that the formation of the array uniformly influences both polarizations, as evidenced in Fig.~\ref{fig: dualpol array}(d).

In conclusion to this section, we have decided on the required shape for the antenna element: It must be two dipoles orthogonal to each other and should be positioned diagonally concerning the formation axis of the sub-arrays.

%% file: design.tex
\section{Design and Simulation}
\label{s: design}

\subsection{Initial Design}

\begin{figure}[t!]
    \centering
    \includegraphics[width = 0.6\linewidth]{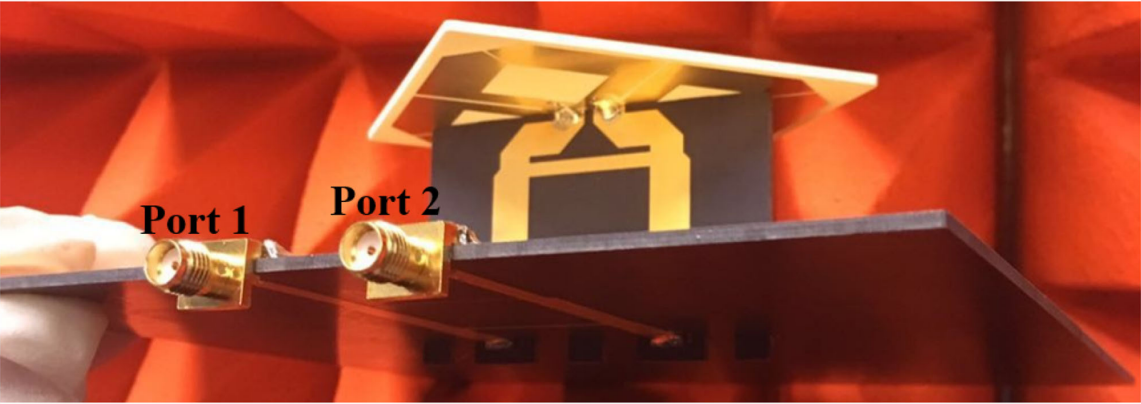}
    \caption{Antenna proposed in~\cite{base station antenna}.}
    \label{fig: start antenna}
\end{figure}

Our starting point for the antenna element was the design proposed in \cite{base station antenna} (depicted in Fig.~\ref{fig: start antenna}), which addresses points (2), (3), (4), and (7) outlined in Table~\ref{table: specs}. Nonetheless, this antenna does not fulfill the other criteria outlined in the table, which are crucial and specific to our application. Utilizing the design proposed in \cite{base station antenna} as a foundation, subsequent steps include (i) adjusting the model for operation within the desired 1200--1800~MHz range, (ii) miniaturizing the antenna to facilitate array formation without generating phase ambiguities, (iii) configuring the three sub-arrays to shape the radiation pattern and achieve a beam tailored to the Milky Way, and finally, (iv) optimizing the ground plane distance for efficient operation in the 1200--1800~MHz band. This section details each of these steps (i-iv) comprehensively.

\begin{figure}[t!]
    \centering
    \begin{subfigure}{0.49\linewidth}
         \centering
         \includegraphics[width = 1\linewidth]{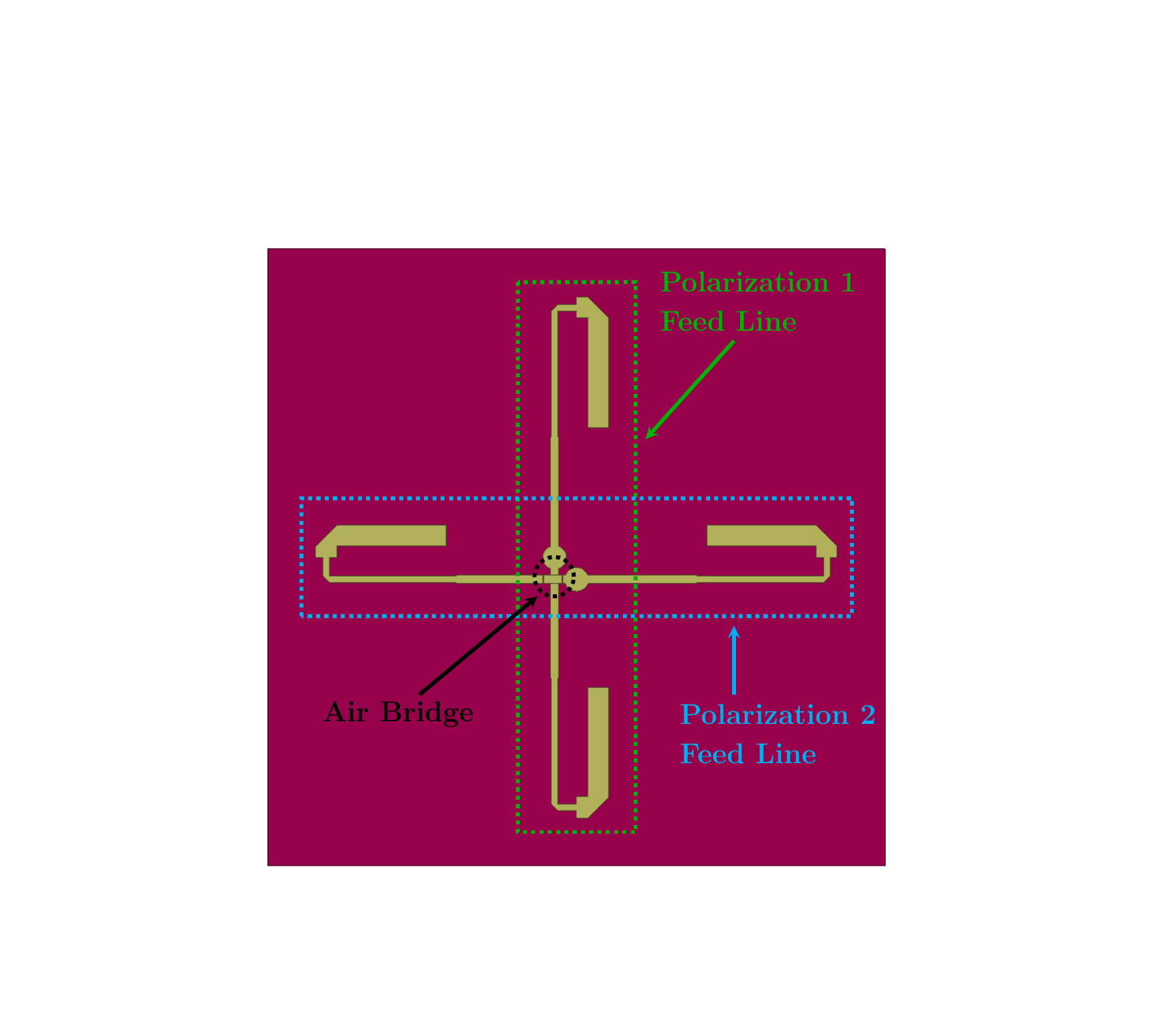}
         \caption{}
     \end{subfigure}
     \begin{subfigure}{0.49\linewidth}
         \centering
         \includegraphics[width = 1\linewidth]{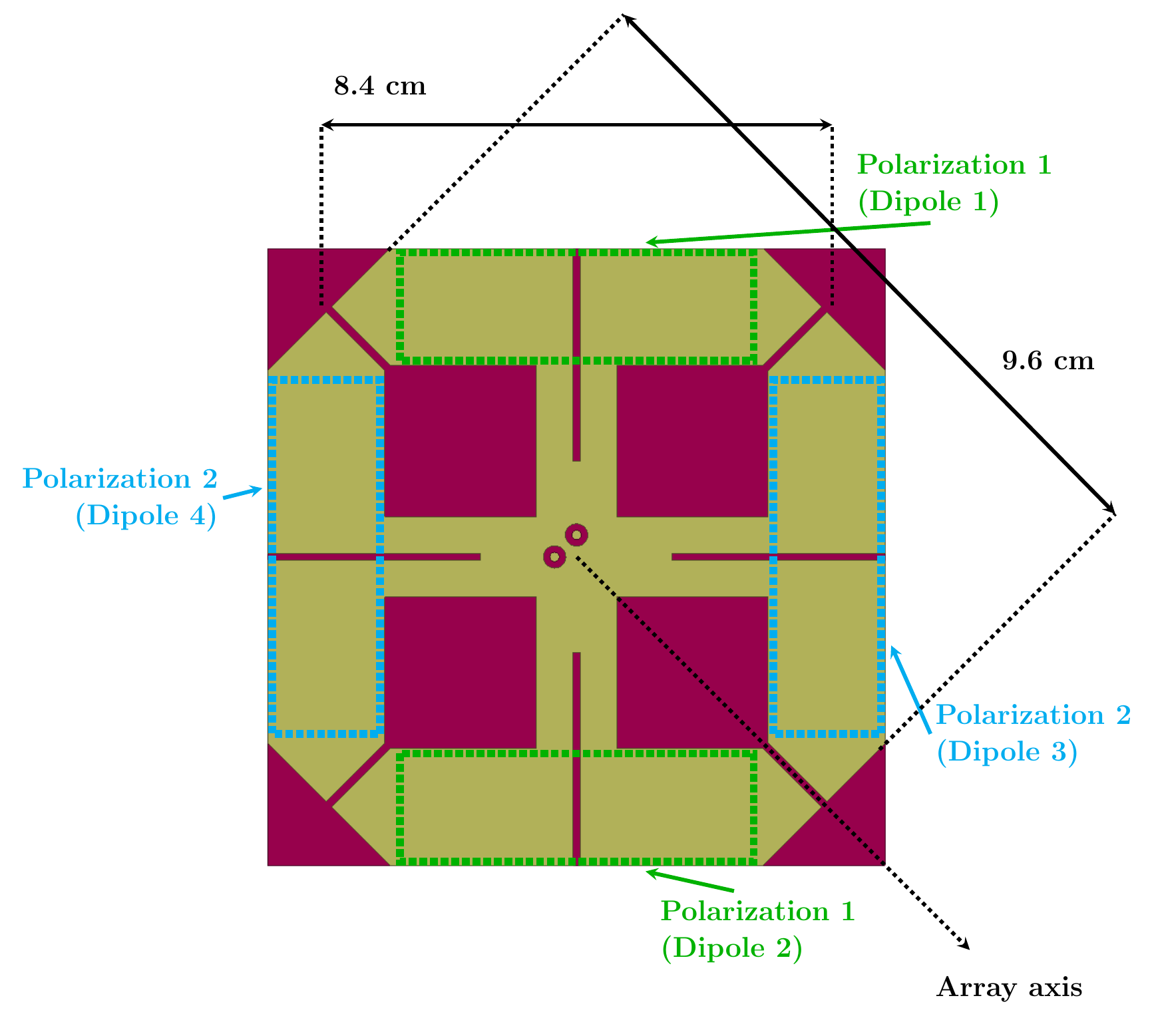}
         \caption{}
     \end{subfigure}
    \caption{HFSS model of the antenna proposed in~\cite{base station antenna}. The antenna is printed on a RO4003C substrate with 1.5~mm thickness and has a size of 8.4$\times$8.4 cm$^2$ when escalated to operate at 1200--1800~MHz. (a) Top Layer. (b) Bottom Layer.}
    \label{fig: hfss starting}
\end{figure}

To modify the antenna element shown in Fig.~\ref{fig: start antenna}, it is imperative to comprehend this design, which is explained in the following paragraphs. The antenna element was modeled in HFSS~\footnote{The High-Frequency Structure Simulator, or HFSS, is a commercial finite element method solver for electromagnetic structures.} and is depicted in Fig.~\ref{fig: hfss starting}.

The antenna comprises four dipoles and two feed lines, each corresponding to one polarization. Dipoles 1 and 2 (panel (b) of the figure) run parallel to each other, associating with the same polarization (referred to as \textit{polarization 1}). Similarly, dipoles 3 and 4 are parallel to each other and orthogonal to dipoles 1 and 2, representing the orthogonal polarization (referred to as \textit{polarization 2}). Since the two feed lines overlap in the top layer, the design proposed in~\cite{base station antenna} incorporates an air-bridge to address this issue.

The question may arise as to why four dipoles are necessary instead of two (e.g., dipoles 1 and 3, which already fulfill the dual-pol requirement). The answer lies in the individual characteristics of dipoles—high reactances and low bandwidth. To mitigate these issues and widen the bandwidth for compatibility with ARTE, the unconventional matching lines depicted in Fig.~\ref{fig: hfss starting}(a)~\cite{base station antenna} are employed. Although this adaptation successfully eliminates the high reactive values, it elevates the real impedance value above 100~$\Omega$. Consequently, two dipoles are arranged in parallel for each polarization to achieve an equivalent impedance of 50~$\Omega$ at each input port.

\subsection{Miniaturization}

If we use the antenna proposed in~\cite{base station antenna} and scale its parameters to operate within the ARTE operational range (1200--1800MHz), we obtain an antenna with a size of 9.6~cm in the direction relevant for forming the array (this antenna is the one shown in Fig.~\ref{fig: hfss starting}). As discussed in Section~\ref{ss: separation}, this is unacceptable since we require a size smaller than 8.3~cm. Therefore, we need to proceed with miniaturizing the antenna.

\begin{figure}[t!]
    \centering
    \begin{subfigure}[B]{0.4\linewidth}
         \centering
         \includegraphics[width = 1\linewidth]{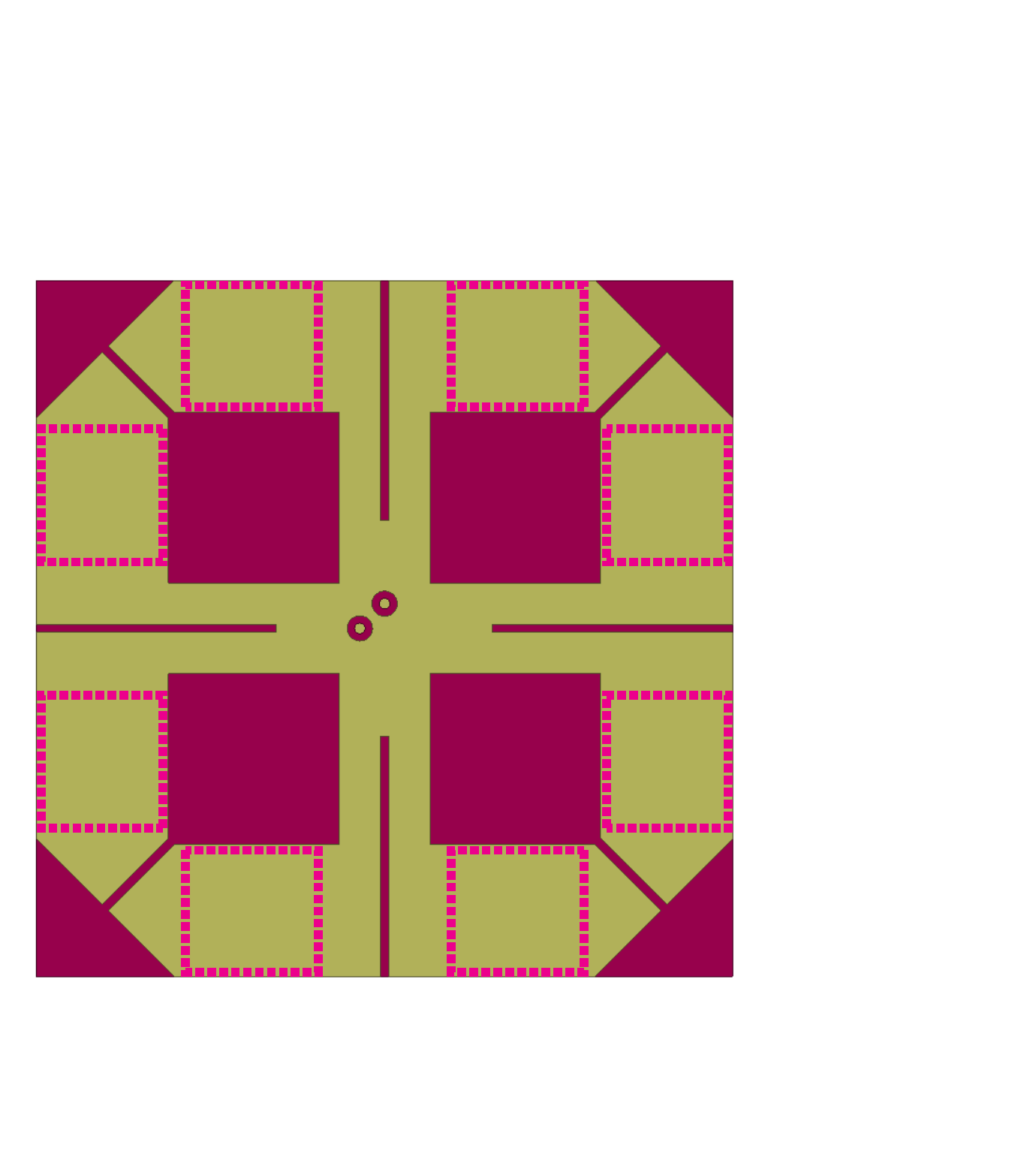}
         \caption{}
     \end{subfigure}
     \begin{subfigure}[B]{0.4\linewidth}
         \centering
         \includegraphics[width = 1\linewidth]{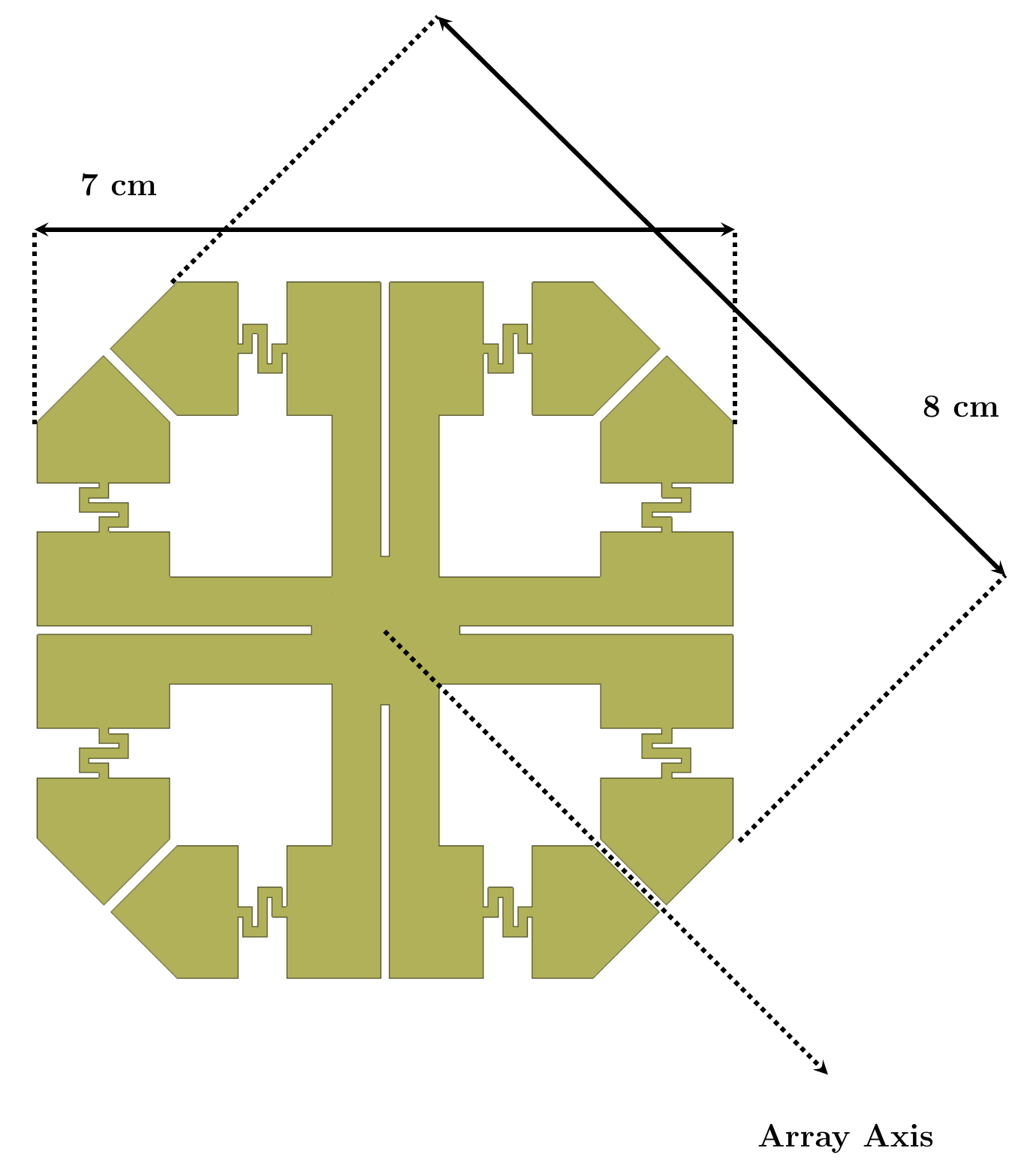}
         \caption{}
     \end{subfigure}
     \begin{subfigure}[B]{0.18\linewidth}
         \centering
         \includegraphics[width = 1\linewidth]{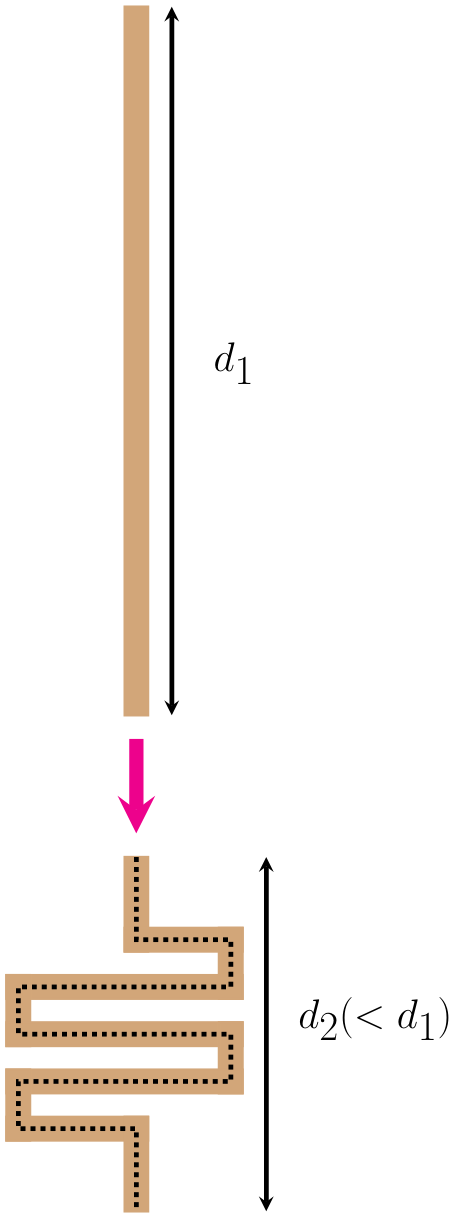}
         \caption{}
     \end{subfigure}
    \caption{Miniaturization of the antenna. (a) Original configuration. The sections requiring modification to achieve miniaturization are highlighted with dashed lines. (b) Miniaturized configuration. (c) Miniaturization of a longitudinal segment.}
    \label{fig: modify sections}
\end{figure}

To effectively miniaturize the antenna, the longitudinal sections of the dipoles must be altered, as depicted in Fig.~\ref{fig: modify sections}(a). All dipoles must be changed in the same manner (both parallel dipoles to avoid asymmetries that could distort the radiation pattern and crossed dipoles since both polarizations must meet the exact requirements).

To achieve miniaturization, we cannot merely scale the dimensions of the dipoles, as this would alter the antenna's operating frequency. To effectively miniaturize a dipole (i.e., reducing its physical dimension without changing its operating frequency), we must reduce its physical length while maintaining its electrical length. In other words, the current traveling through the dipoles must experience the same phase variation over a shorter physical distance. There are numerous techniques for miniaturizing antennas, including incorporating holes in conductors~\cite{small antennas}, utilizing materials with higher permittivity or permeability, integrating metamaterials~\cite{zor antenna}, or employing the twisting of longitudinal structures~\cite{spiral antenna}. As the dipole is a longitudinal structure that can be readily twisted, we have opted for that miniaturization technique in this work, as illustrated in Fig.~\ref{fig: modify sections}(c). Consequently, we achieve the miniaturization shown in Fig.~\ref{fig: modify sections}(b). Observing the size reduction from 9.6~cm to 8~cm, we conclude that the antenna element is feasible for operation in an array.

\subsection{Final Design of the Antenna Element}

\begin{figure}[t!]
    \centering
    \begin{subfigure}[t!]{0.4\linewidth}
         \centering
         \includegraphics[height = 0.75\linewidth]{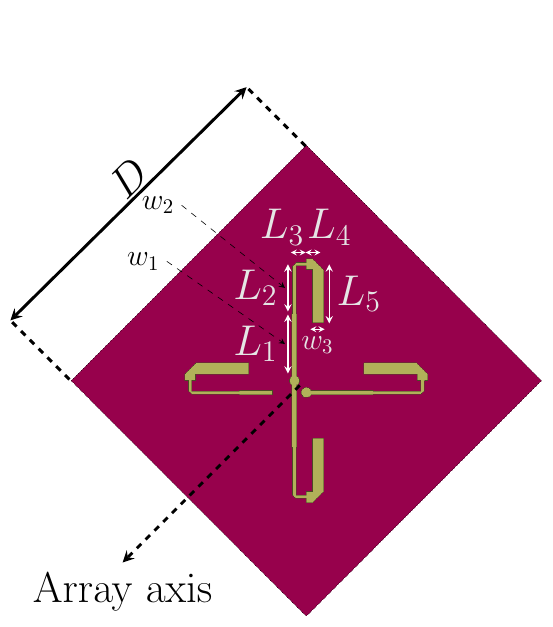}
         \caption{}
     \end{subfigure}
     \begin{subfigure}[t!]{0.4\linewidth}
         \centering
         \includegraphics[height = 0.75\linewidth]{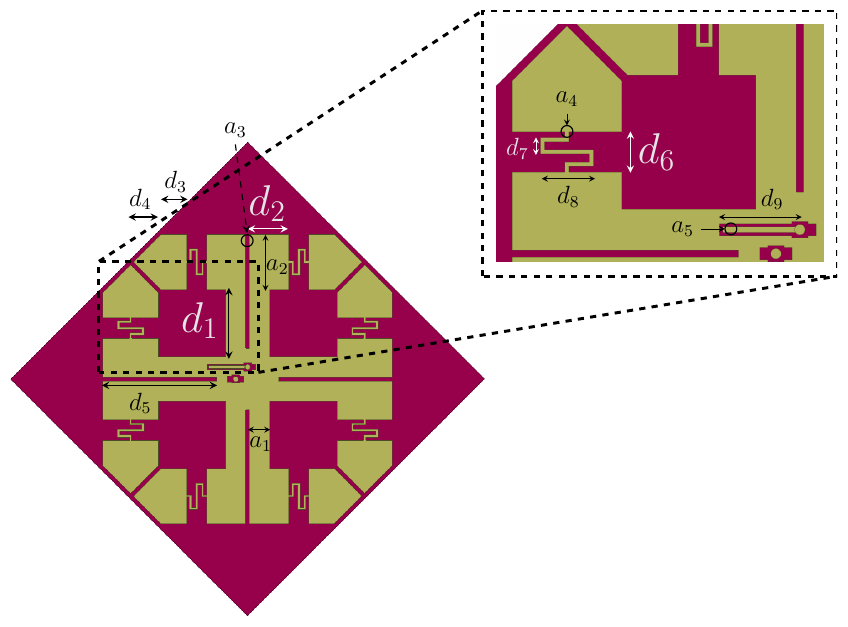}
         \caption{}
     \end{subfigure}
     \begin{subfigure}[t!]{0.34\linewidth}
         \centering
         \includegraphics[height = 0.5\linewidth]{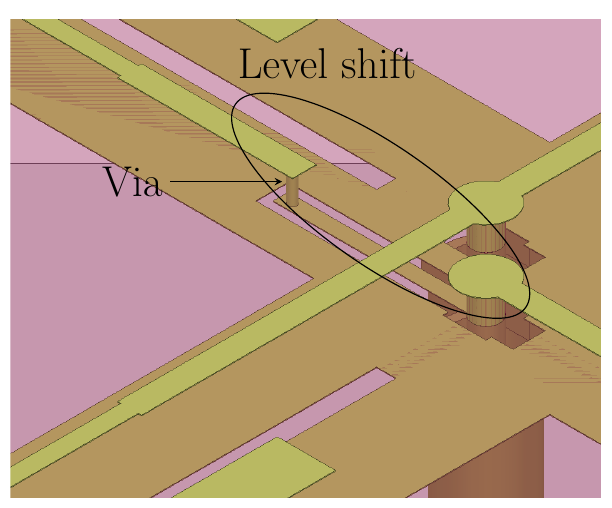}
         \caption{}
     \end{subfigure}
     \begin{subfigure}[t!]{0.3\linewidth}
         \centering
         \includegraphics[height = 0.75\linewidth]{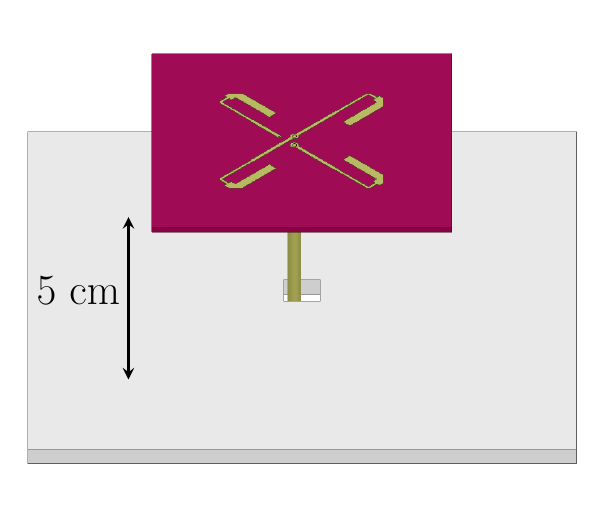}
         \caption{}
     \end{subfigure}
    \caption{HFSS model of the final design of the antenna element. (a) Top Layer. (b) Bottom Layer. (c) Level shift to address the crossover between feed lines. (d) Incorporation of the reflecting plane.}
    \label{fig: final individual design}
\end{figure}

After adapting the antenna design proposed in~\cite{base station antenna} to operate within the 1200--1800~MHz range and subsequently miniaturizing it for array formation feasibility, we arrived at the configuration depicted in Fig.~\ref{fig: final individual design}. For this design, we utilized a 1.5~mm thick RO4003C substrate (dielectric permittivity of 3.55 and loss tangent of 0.0027). Dimensions are tabulated in Table~\ref{table: params}.

The finalized printed antenna design encompasses four dipoles paired with two feed lines, catering to each polarization. Ideally, both polarizations' feeds would be centrally located, which poses challenges due to coaxial cable overlap. To circumvent this, we made minimal adjustments to the feed points and transitioned from SMA to more compact IPEX connectors. Furthermore, Fig.~\ref{fig: final individual design}(a) underscores a crossover between the upper layer feed lines because of the orthogonal nature of both polarizations. This crossover was addressed using the level shift illustrated in Fig.~\ref{fig: final individual design}(c).

Regarding the tolerance in the antenna dimensions, a sensibility analysis was conducted in ANSYS HFSS considering three different values for each of the parameters: $d_2$ (9.5~mm, 9.55~mm, 9.6~mm), $L_5$ (14.4~mm, 14.45~mm, 14.5~mm), and $L_1$ (16.45~mm, 16.5~mm, 16.55~mm), resulting in a total of 27 simulations. The objective was to assess the impact of potential variations in the manufacturing process on the reflections observed at each polarization port, as depicted in Fig. 12. Notably, the variations were negligible when the parameters varied within 0.1 mm. The most significant deviation observed was from $-$15.5 dB to $-$17.5 dB, equivalent to a variation of 0.01 on a linear scale. This suggests that a precision greater than 0.1 mm is unnecessary, a tolerance achievable by many printed circuit-making machines.

Finally, it is worth remembering that ARTE requires a Front-to-Back Ratio greater than 10~dB. As mentioned at the beginning of this section, the antenna proposed in~\cite{base station antenna} already meets this criterion. The way the antenna meets the criterion is fairly standard: a reflecting plane is used at the back of the antenna. The distance between the antenna and the reflecting plane is relevant, and it is known that the ideal is to place the reflector at a distance of $\lambda_c / 4$, with $\lambda_c$ being the wavelength of the central operating frequency. Since ARTE operates between 1200 and 1800~MHz, the center frequency is 1500~MHz. Therefore, the reflecting plane must be positioned at a distance of 5~cm from the antenna, as shown in Fig.~\ref{fig: final individual design}(d). It should be noted that all the antenna parameters, shown in Table~\ref{table: params}, were optimized considering the presence of the reflecting plane.

\begin{table}[t!]
    \caption{Physical dimensions of the antenna element. The values shown were obtained in the simulation optimization process, but a precision of two decimal points is unnecessary.}
    \centering
    \scalebox{0.9}[0.9]{%
    \begin{tabular}{c|c||c|c}
    \hline \hline
    \multicolumn{2}{c||}{\textbf{Top Layer}} & \multicolumn{2}{c}{\textbf{Bottom Layer}} \\ \hline
    \textbf{Parameter} & \textbf{Value (mm)} & \textbf{Parameter} & \textbf{Value (mm)} \\ \hline
    $D$ & 8.2  & $d_1$ & 16.46 \\ 
    $L_1$ & 16.48 & $d_2$ & 9.55 \\ 
    $L_2$ & 12.19  & $d_3$ & 6.2 \\ 
    $L_3$ & 2.95  & $d_4$ & 6.75 \\ 
    $L_4$ & 2.95  & $d_5$ & 27.85 \\   
    $L_5$ & 14.45  & $d_6$ & 5 \\ 
    $w_1$ & 1.1 & $d_7$ & 2 \\ 
    $w_2$ & 0.8 & $d_8$ & 6.5 \\ 
    $w_3$ & 2.7 & $d_9$ & 9.5 \\ 
    - & - & $a_1$ & 5 \\ 
    - & - & $a_2$ & 13.5 \\ 
    - & - & $a_3$ & 0.9 \\ 
    - & - & $a_4$ & 0.5 \\ 
    - & - & $a_5$ & 0.7 \\ \hline \hline 
    \end{tabular}
    }
    \label{table: params}
\end{table} 

\begin{figure}[t!]
    \centering
    \begin{subfigure}{0.49\linewidth}
         \centering
         \includegraphics[width = 1\linewidth]{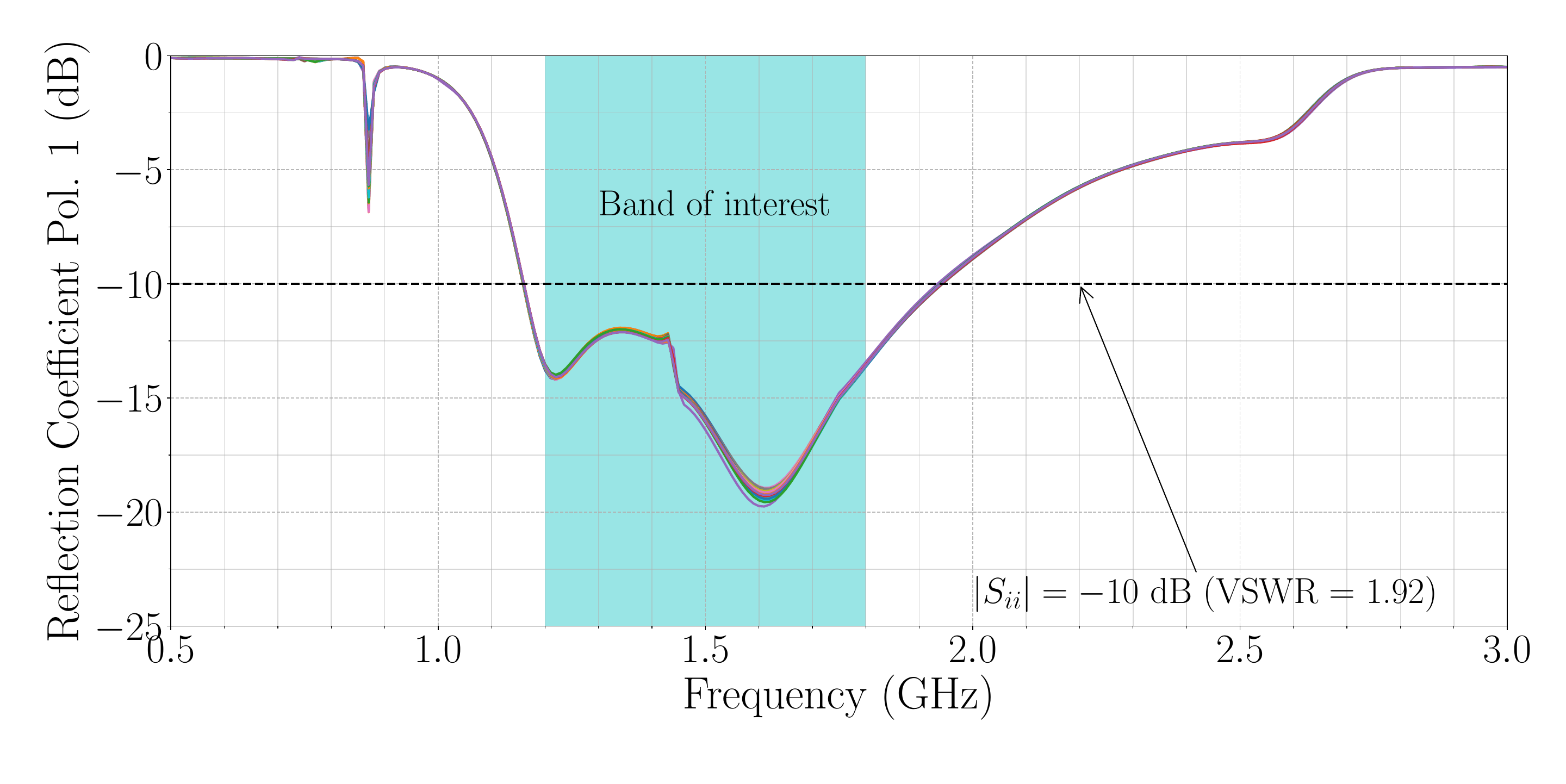}
         \caption{}
     \end{subfigure}
     \begin{subfigure}{0.49\linewidth}
         \centering
         \includegraphics[width = 1\linewidth]{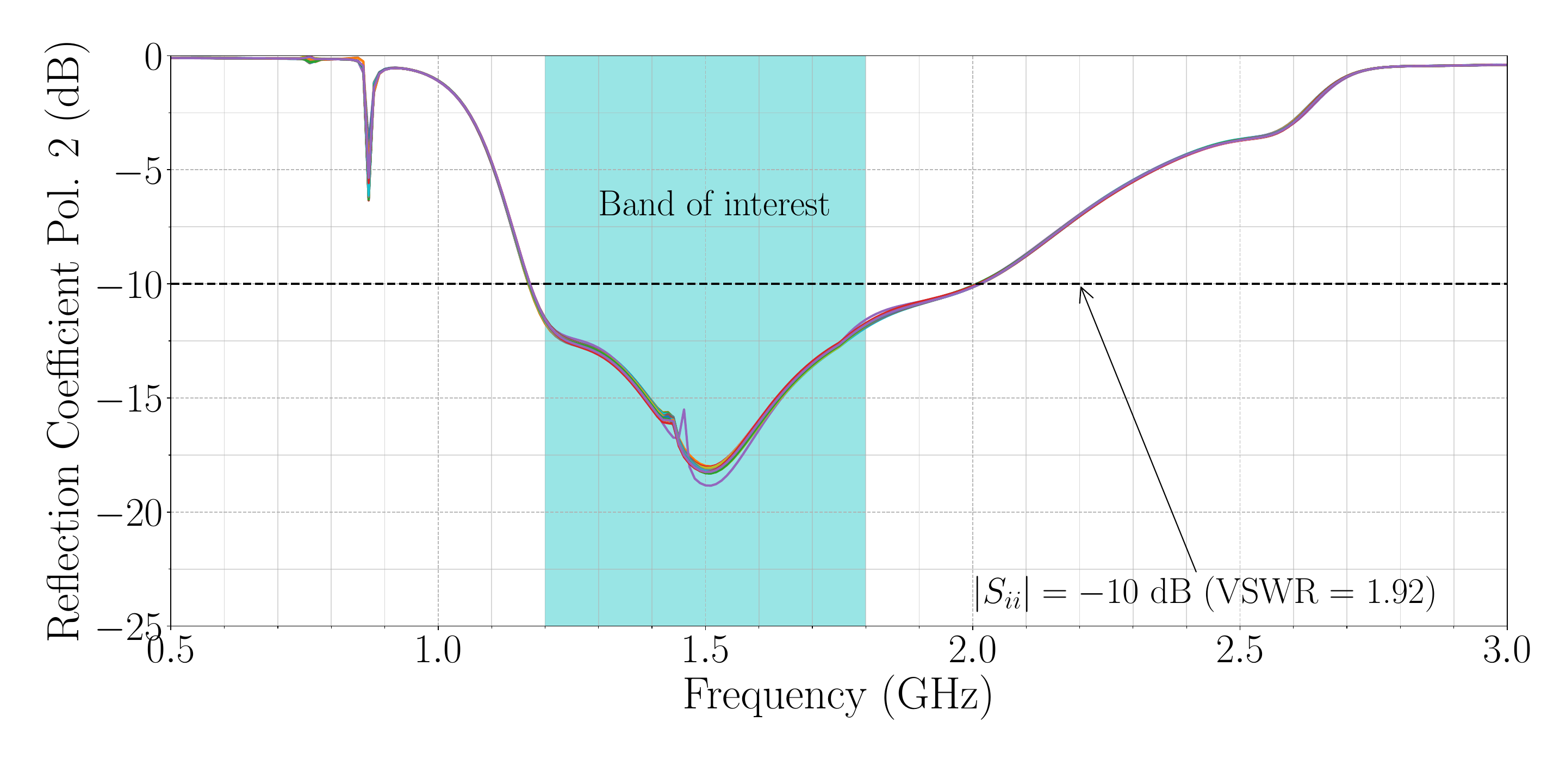}
         \caption{}
     \end{subfigure}
    \caption{Sensibility simulations of the reflection coefficient at each polarization port in the antenna element. Variations of up to 0.1~mm were made in the parameters $d_2$, $L_5$, and $L_1$, leading to 27 simulations. (a) Polarization 1. (b) Polarization 2.}
    \label{fig: reflections sim}
\end{figure}

\subsection{Design of the ARTE Antenna}

\begin{figure}[t!]
    \centering
    \includegraphics[width = 0.5\linewidth]{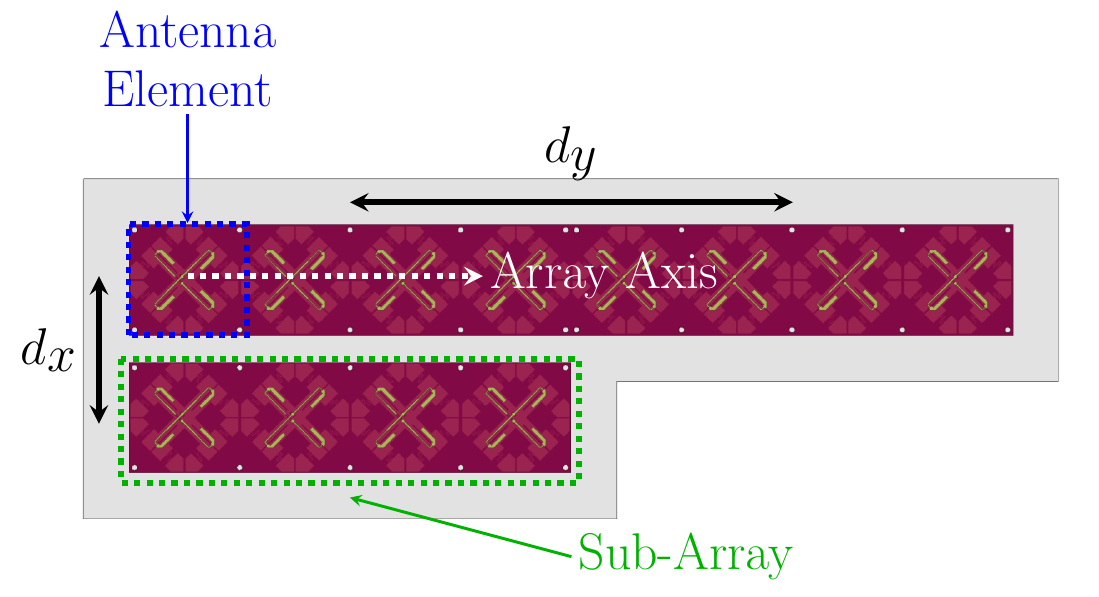}
    \caption{HFSS model of the complete ARTE array.}
    \label{fig: arte hfss}
\end{figure}

After finalizing the antenna element design, the subsequent step involves designing the three sub-arrays constituting the complete ARTE array. The antenna element is replicated four times along the array axis to construct each sub-array, as shown in Fig.~\ref{fig: arte hfss}. Then, three sub-arrays are placed in an L-shape to form the entire ARTE antenna, as mentioned throughout the paper. As stated in Section~\ref{ss: separation}, the sub-arrays should be separated by at most 8.3~cm in the $x$-axis and 32.2~cm in the $y$-axis. However, the antenna aims to be placed in a dome partially covering the emission from the horizon (Fig.~\ref{fig: dome}). Therefore, if we follow the development of Section~\ref{ss: separation}, we see that for the wide part of the radiation pattern (the XZ cut), we must consider $\Theta_0 < 90^\circ$, which implies that the distance $d_x$ can be greater than $\lambda/2$. Specifically, if we consider $\Theta_0 = 60^\circ$, the criterion for $d_x$ changes to $d_x>$ 9.6~cm. In conclusion, depending on the dome covering, there is some freedom when choosing the distance $d_x$ (it could be greater than 8.3~cm).

Simulations of efficiency, radiation pattern, and reflection coefficient $S_{11}$ at each of the 24 ports (three sub-arrays, each with four antenna elements and two polarizations per antenna) were done with ANSYS HFSS. The reflections for $d_x = $ 8.2~cm and $d_x = $ 10.2~cm are shown in Figs.~\ref{fig: reflections sim}(a) and (b), respectively. When the separation is $d_x = $ 8.2~cm, some antenna elements do not satisfy the $<-10$~dB reflection criterion. Such antennas are precisely those found in the adjacent sub-arrays on the $x$-axis. The problem arises from the coupling between antennas due to the small separation. Therefore, separating the sub-arrays further on the $x$-axis is the solution, as shown in Fig.~\ref{fig: reflections sim}(b). However, for the initial setup of the ARTE experiment, a separation distance of $d_x = $ 8.2~cm was chosen since this was the original criteria established in Section~\ref{ss: separation}, which accounts for the capacity of the antenna on its own, without its incorporation into a dome.

\begin{figure}[t!]
    \centering
    \begin{subfigure}{0.49\linewidth}
         \centering
         \includegraphics[width = 1\linewidth]{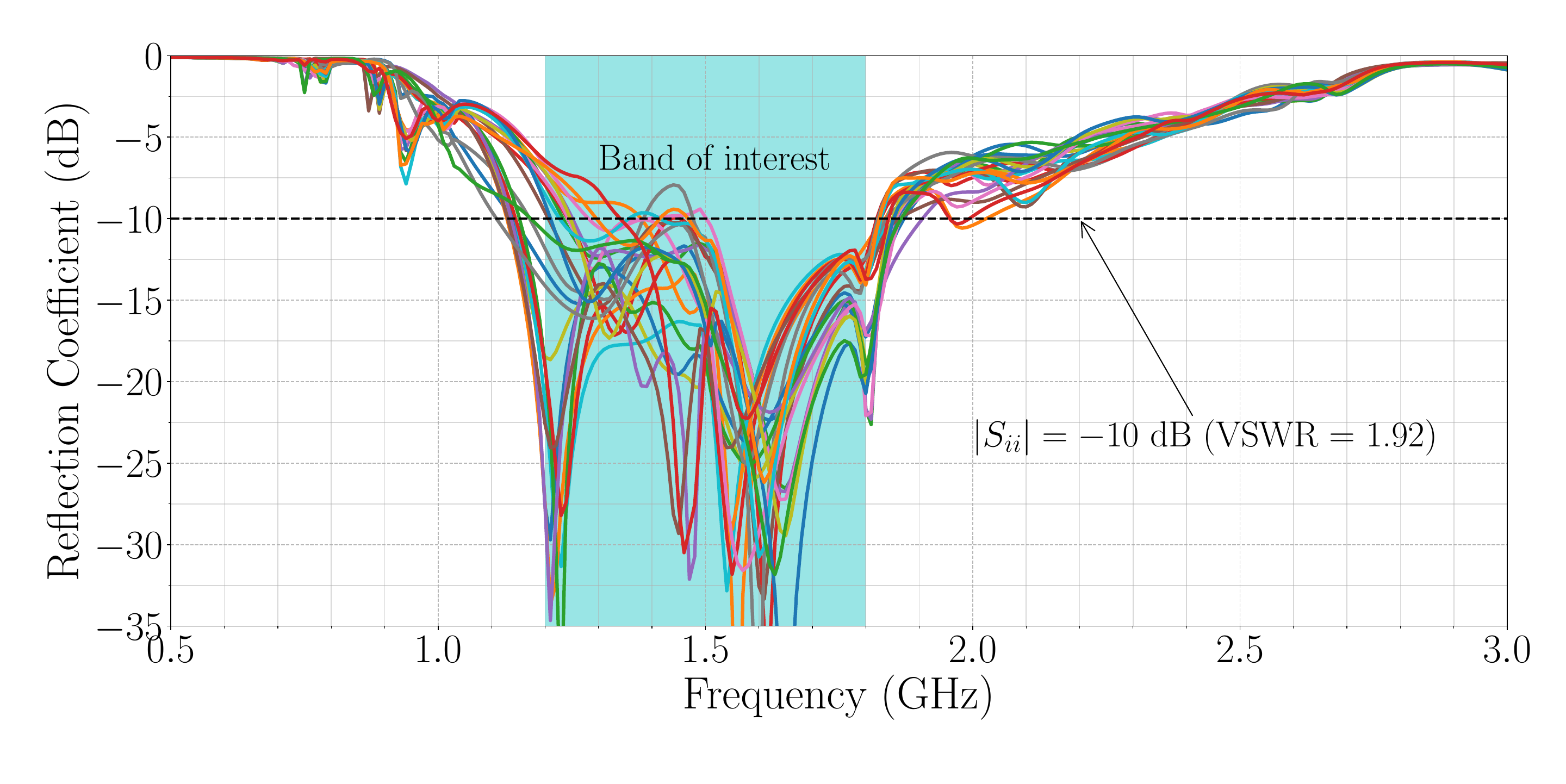}
         \caption{}
     \end{subfigure}
     \begin{subfigure}{0.49\linewidth}
         \centering
         \includegraphics[width = 1\linewidth]{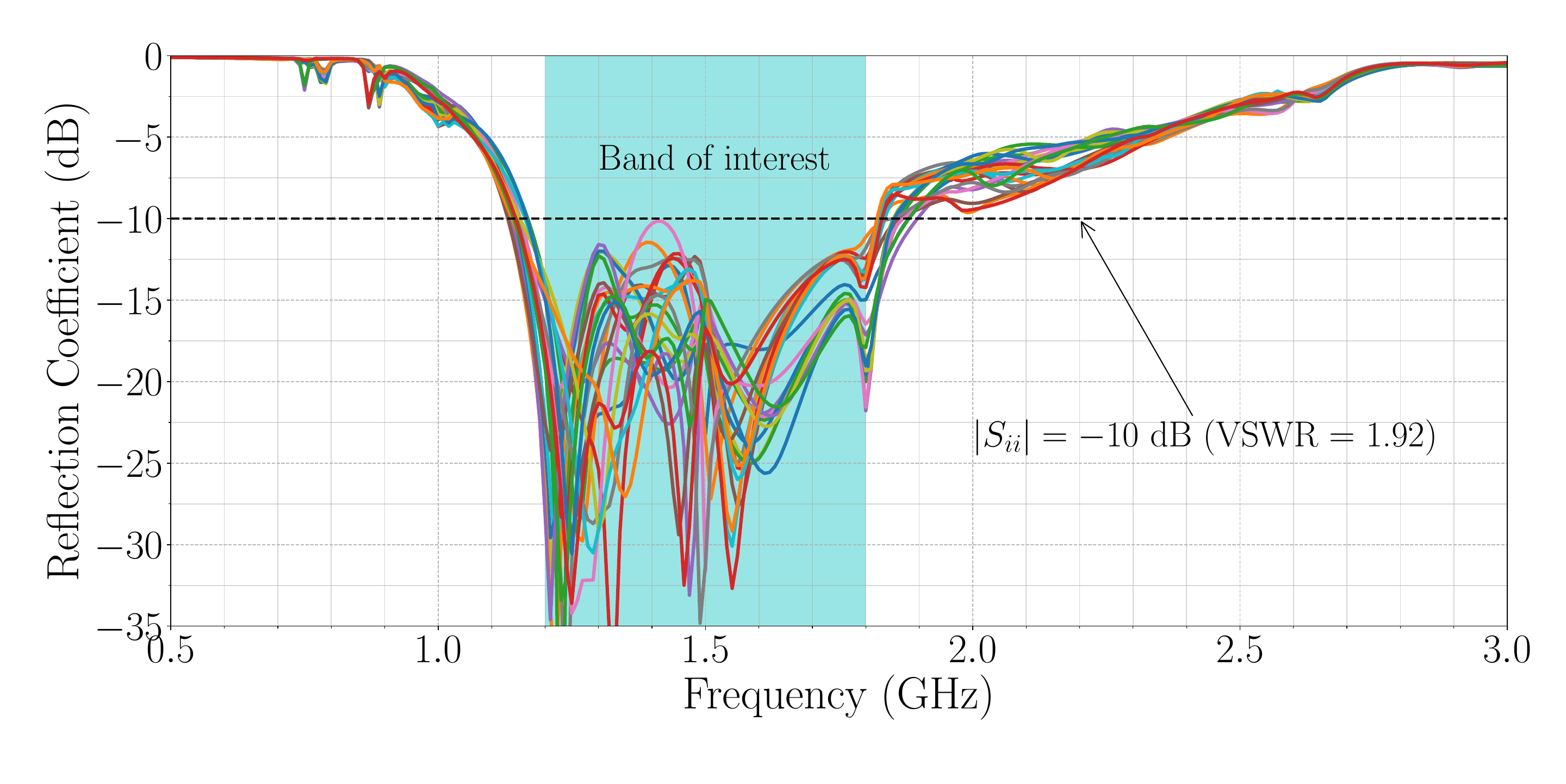}
         \caption{}
     \end{subfigure}
    \caption{Simulation of the reflection coefficient at each of the 24 ports. (a) $d_x = $ 8.2~cm. (b) $d_x = $ 10.2~cm.}
    \label{fig: reflections sim}
\end{figure}

\begin{figure}[t!]
    \centering
    \includegraphics[width = 0.5\linewidth]{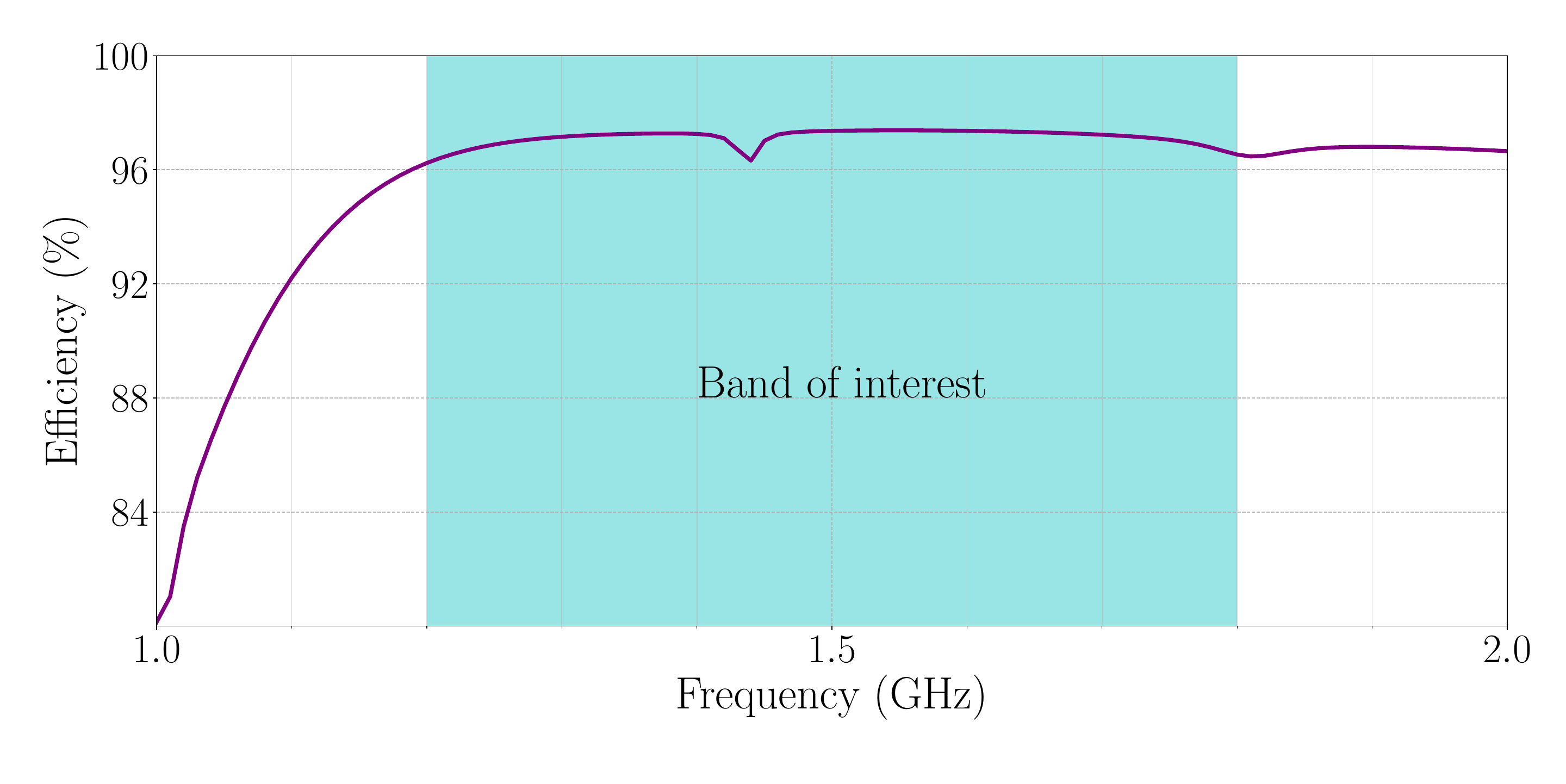}
    \caption{Simulated efficiency of a single element.}
    \label{fig: efficiency}
\end{figure}

Finally, the simulated radiation efficiency is shown in Fig.~\ref{fig: efficiency} for a single element. The losses remain small given that the selected substrate possesses a low loss tangent of 0.0027 and the antenna doesn't employ a traveling wave that propagates through an extended structure. This results in a high radiation efficiency that surpasses the criterion of $> 90$\%.

%% file: meas.tex
\section{Measurements and Discussion}
\label{s: meas}

\subsection{Fabrication and Measurement Setup}

\begin{figure}[t!]
    \centering
    \begin{subfigure}[b]{0.352\linewidth}
         \centering
         \includegraphics[width = 0.85\linewidth, height = 1.105\linewidth]{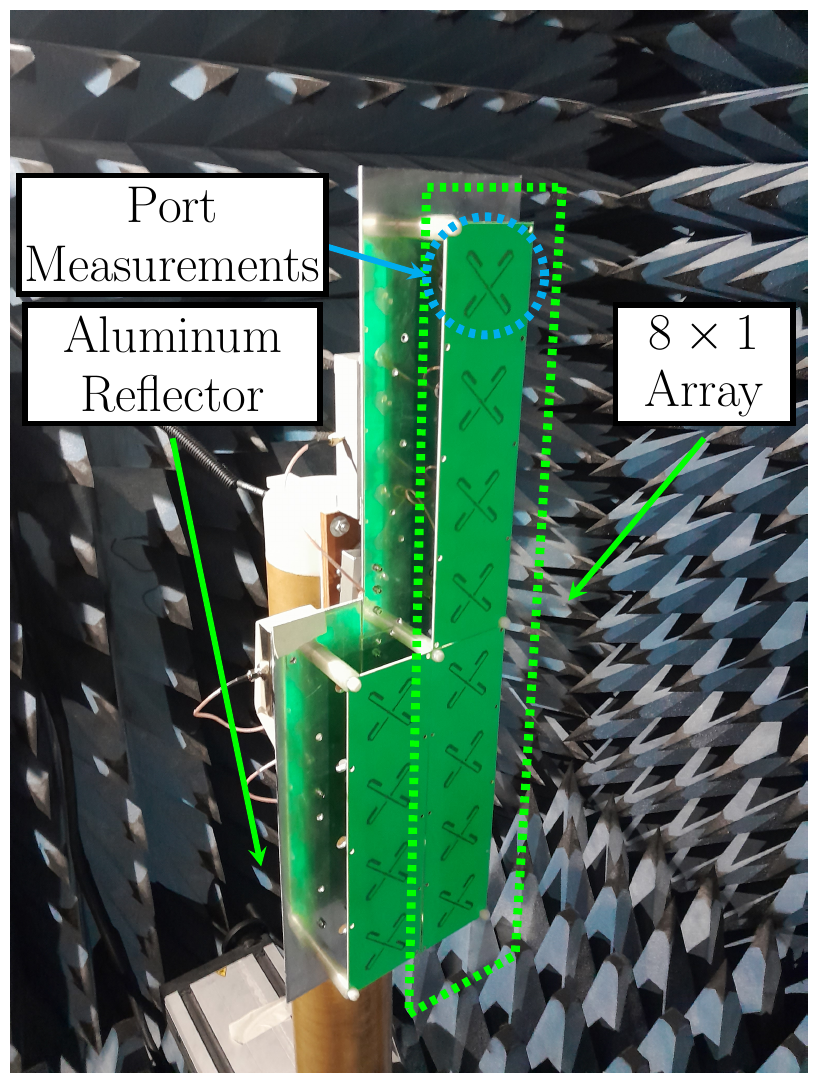}
         \caption{}
     \end{subfigure}
     \begin{subfigure}[b]{0.2368\linewidth}
         \centering
         \includegraphics[width = 0.85\linewidth, height = 1.643\linewidth]{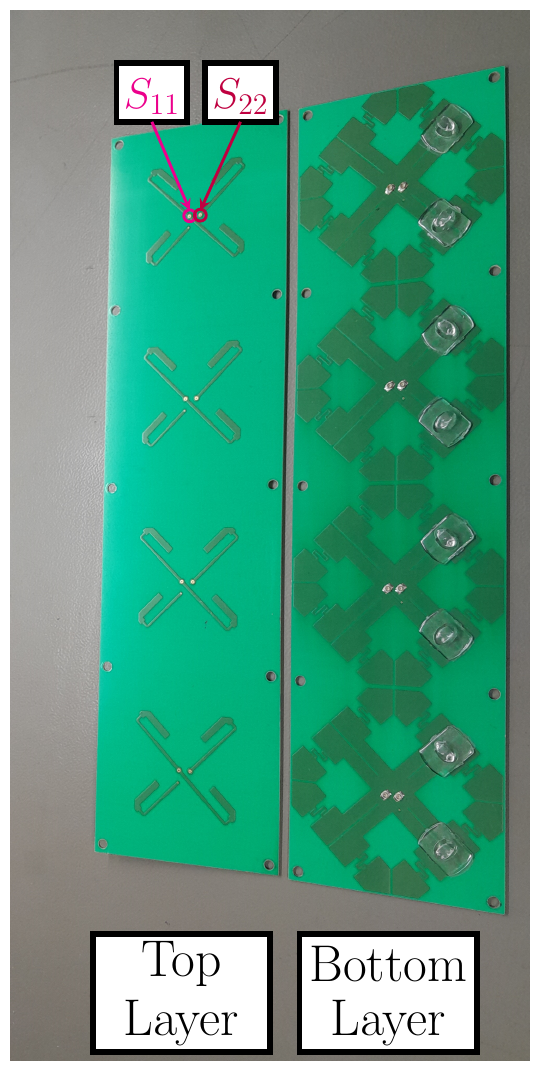}
         \caption{}
     \end{subfigure}
    \caption{(a) Fabricated prototype of the L-shaped array. (b) Top and Bottom layers.}
    \label{fig: fabricated array}
\end{figure}

To validate the theoretical structure and the predicted performance, we constructed and tested the L-shaped array shown in Fig.~\ref{fig: fabricated array}(a). The aluminum reflector and the sub-arrays are mechanically assembled with PLA (Polylactic acid) supports and nylon screws. Scattering parameter measurements were made with a vector network analyzer, while radiation pattern measurements were made in an anechoic chamber. 

\subsection{Port Parameters}

\begin{figure}[t!]
    \centering
    \begin{subfigure}[b]{0.48\linewidth}
         \centering
         \includegraphics[width = 1\linewidth]{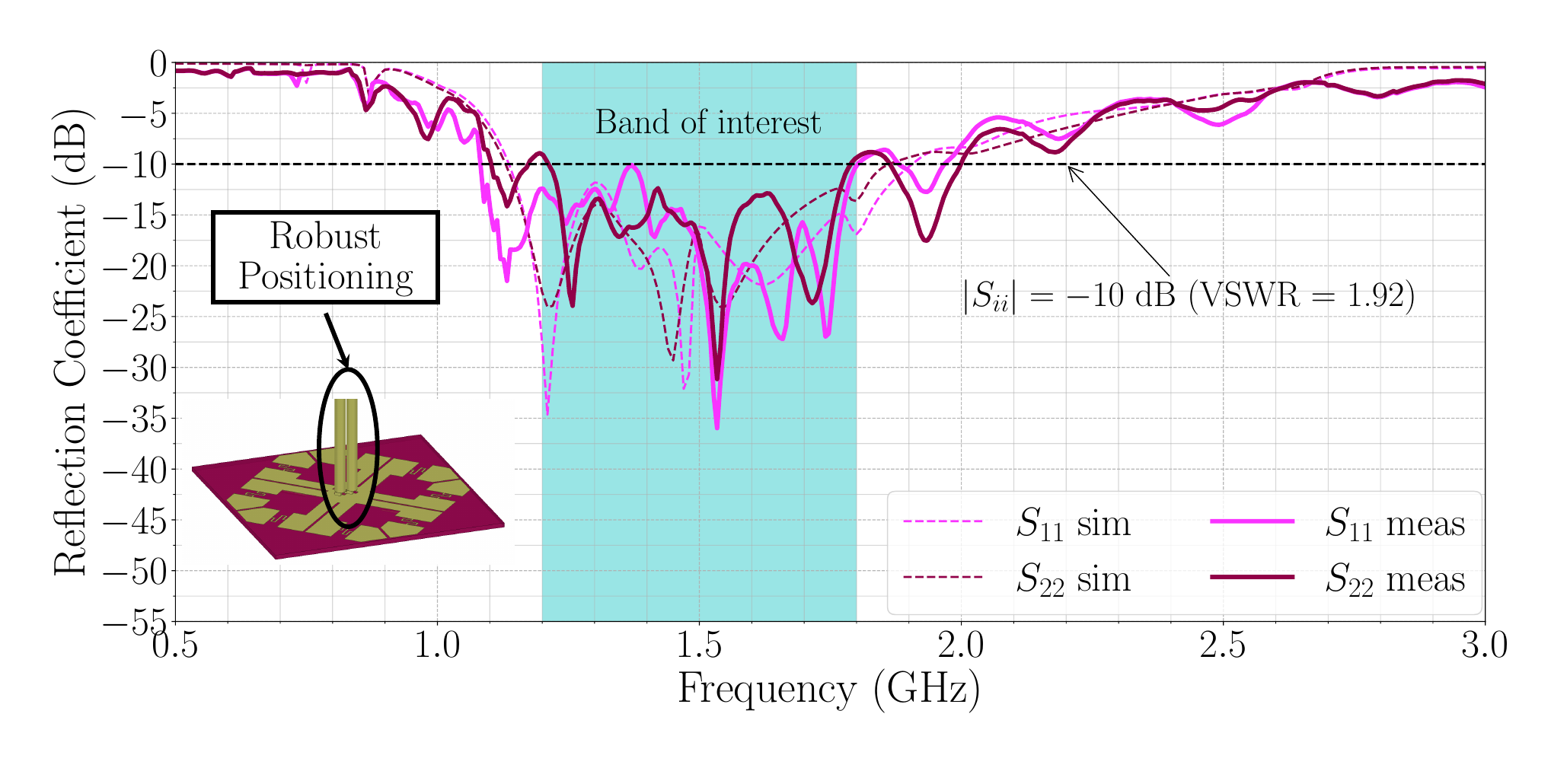}
         \caption{}
     \end{subfigure}
     \begin{subfigure}[b]{0.48\linewidth}
         \centering
         \includegraphics[width = 1\linewidth]{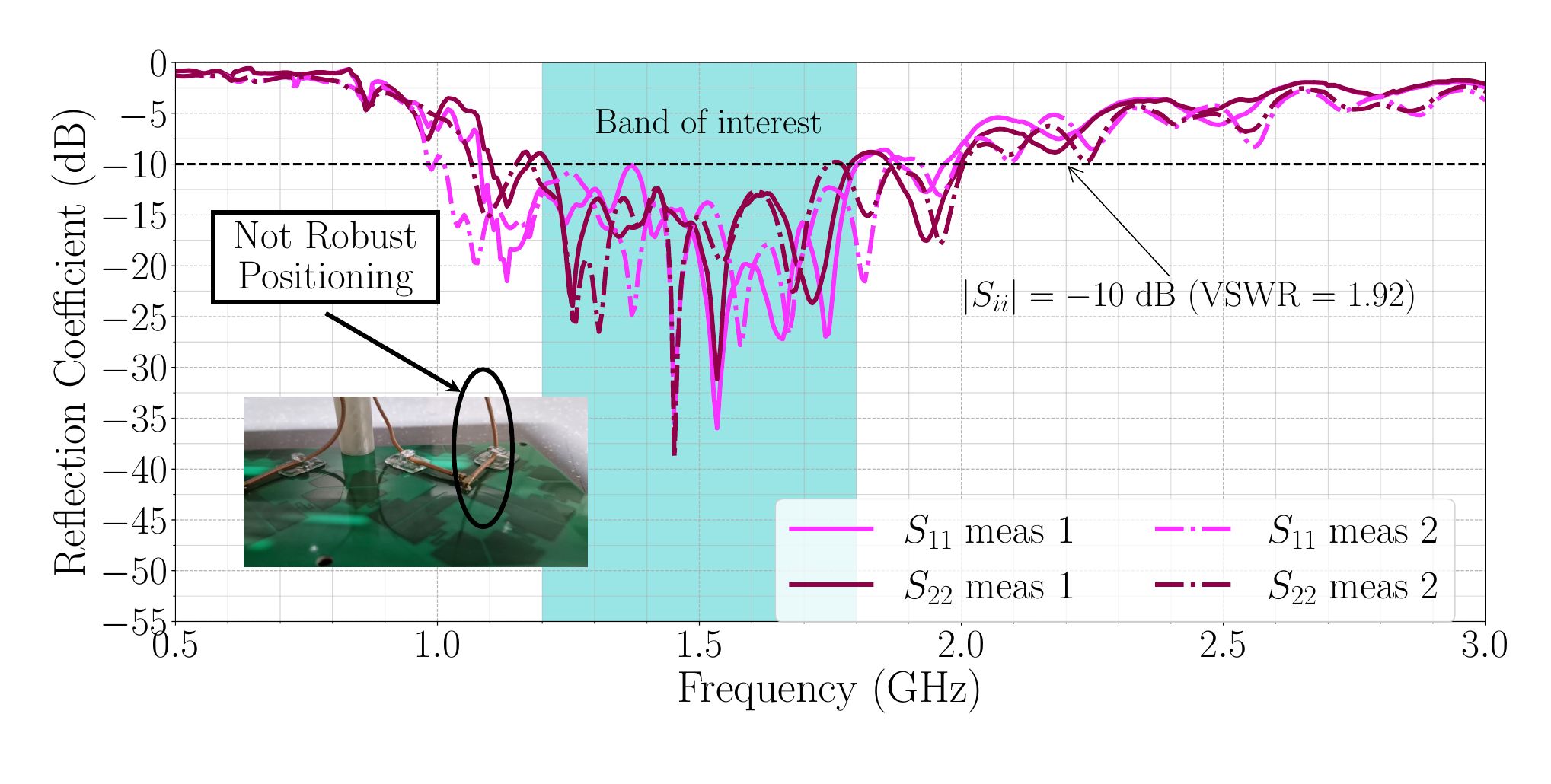}
         \caption{}
     \end{subfigure}
     \begin{subfigure}[b]{0.48\linewidth}
         \centering
         \includegraphics[width = 1\linewidth]{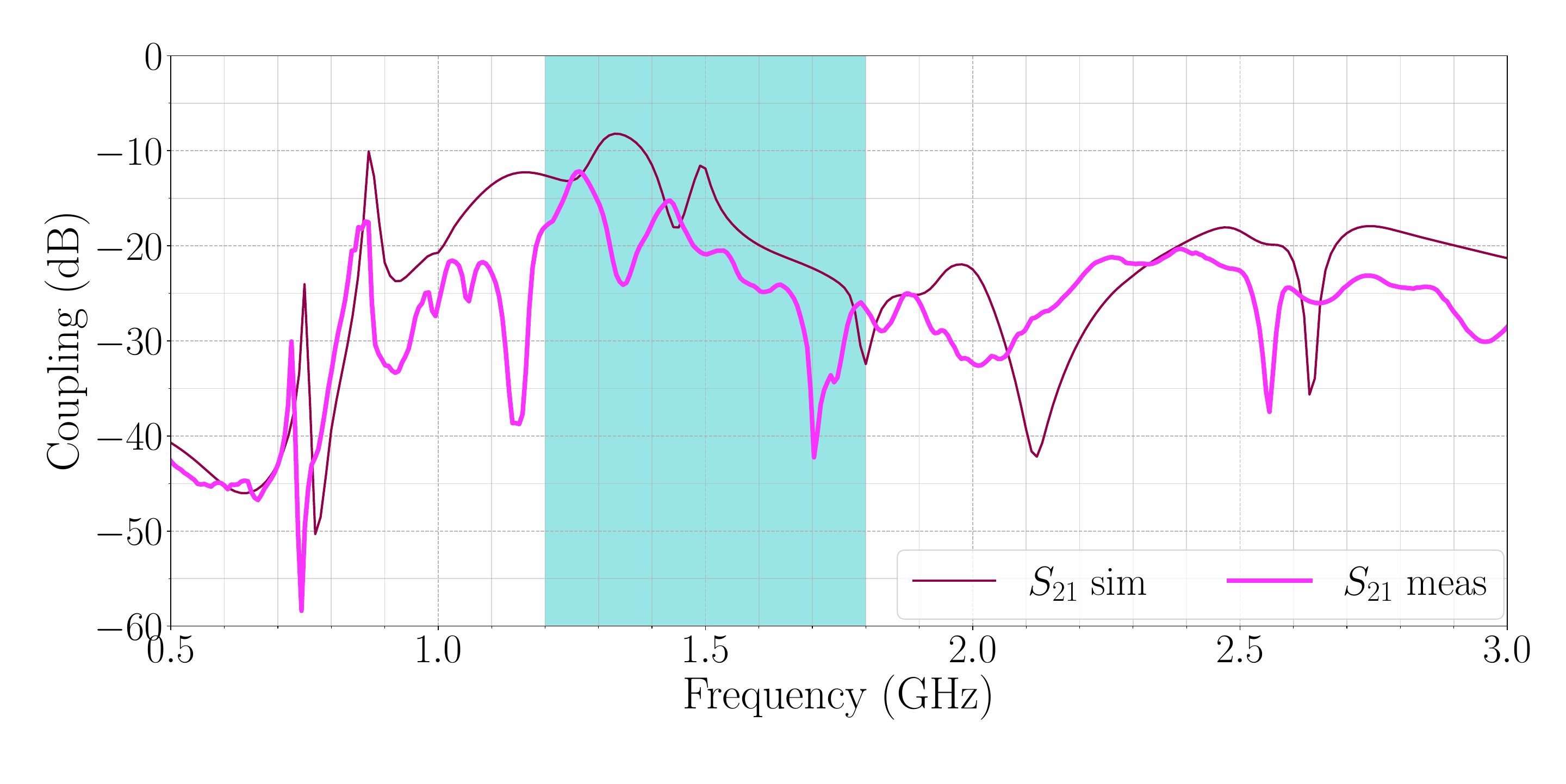}
         \caption{}
     \end{subfigure}
    \caption{(a) Simulated and measured reflections of a single element. (b) Two consecutive measurements considering different positions of the cables. (c) Simulation and measurement of the coupling between polarizations.}
    \label{fig: meas reflections}
\end{figure}

Fig.~\ref{fig: meas reflections} shows the simulated and measured port parameters. Measurements of interest are the reflections at each polarization port, $S_{11}$ and $S_{22}$, and the couple between polarizations, $S_{21}$ (see Fig.~\ref{fig: fabricated array}(b)). To characterize the port parameters and compare them with the simulations, we chose to measure the two polarizations of a single antenna in the array (shown in Fig.~\ref{fig: fabricated array}(a)) while keeping all the other antennas loaded. As shown in Fig.~\ref{fig: meas reflections}(a), the antenna is matched in the band of interest, agreeing with the obtained simulation results. The general shapes of the $S_{11}$ and $S_{22}$ parameters are consistent between simulations and measurements. However, there are differences concerning the resonance behavior and the spectral features. Firstly, we observe a ripple across the entire frequency range, attributed to the IPEX connector used, which could not be accurately modeled in the simulation due to its complex structure. However, this ripple does not compromise the antenna's performance within the band of interest. Secondly, regarding the resonance behavior, we have observed that the positioning of the cables with the IPEX connectors lacks repeatability, leading to changes in the number of resonances and their frequencies. This lack of repeatability is evident in Fig.~\ref{fig: meas reflections}(b), where two consecutive measurements of the same antenna were conducted while moving the cables. While the reflection coefficients remain within the established criteria, using IPEX connectors introduces unwanted instability. This issue will be improved for the full implementation of the instrument, where we plan to employ rigid printed circuit boards for power distribution, as depicted in the original design in Fig.~\ref{fig: start antenna}. 

The isolation between the two polarization ports is shown in Fig.~\ref{fig: meas reflections}(c). We see an isolation greater than 10~dB in the entire band of interest and greater than 15~dB in 90\% of the band. Once again, a difference can be seen between simulation and measurement, which is attributed to the instability of the IPEX connector. While the isolation is comparatively lower than that of the original design depicted in Fig.~\ref{fig: start antenna} (which achieves an isolation greater than 15~dB across the entire band), it is essential to note that the design presented in this work was miniaturized to enable the formation of an array without phase ambiguities. This miniaturization necessitated the incorporation of meanders, which contributed to cross-polarization effects. As the antenna operates in reception mode, the power coupled to port 2 (pol 2) from port 1 (pol 1) is the power that is reflected in the latter. Even under the worst-case scenario where the antenna exhibits an input return loss of 10~dB, the coupled power from one polarization to another would be determined by $r_{12} = |S_{12}|^2 / (1 - |S_{11}|^2) =$~0.001~\cite{coupling}, which is considered negligible for polarimetry calculations. 

\subsection{Radiation Pattern}

\begin{figure}[t!]
    \centering
    \begin{subfigure}[b]{0.32\linewidth}
         \centering
         \includegraphics[width = 0.85\linewidth, trim = {14cm 0 13cm 0}, clip]{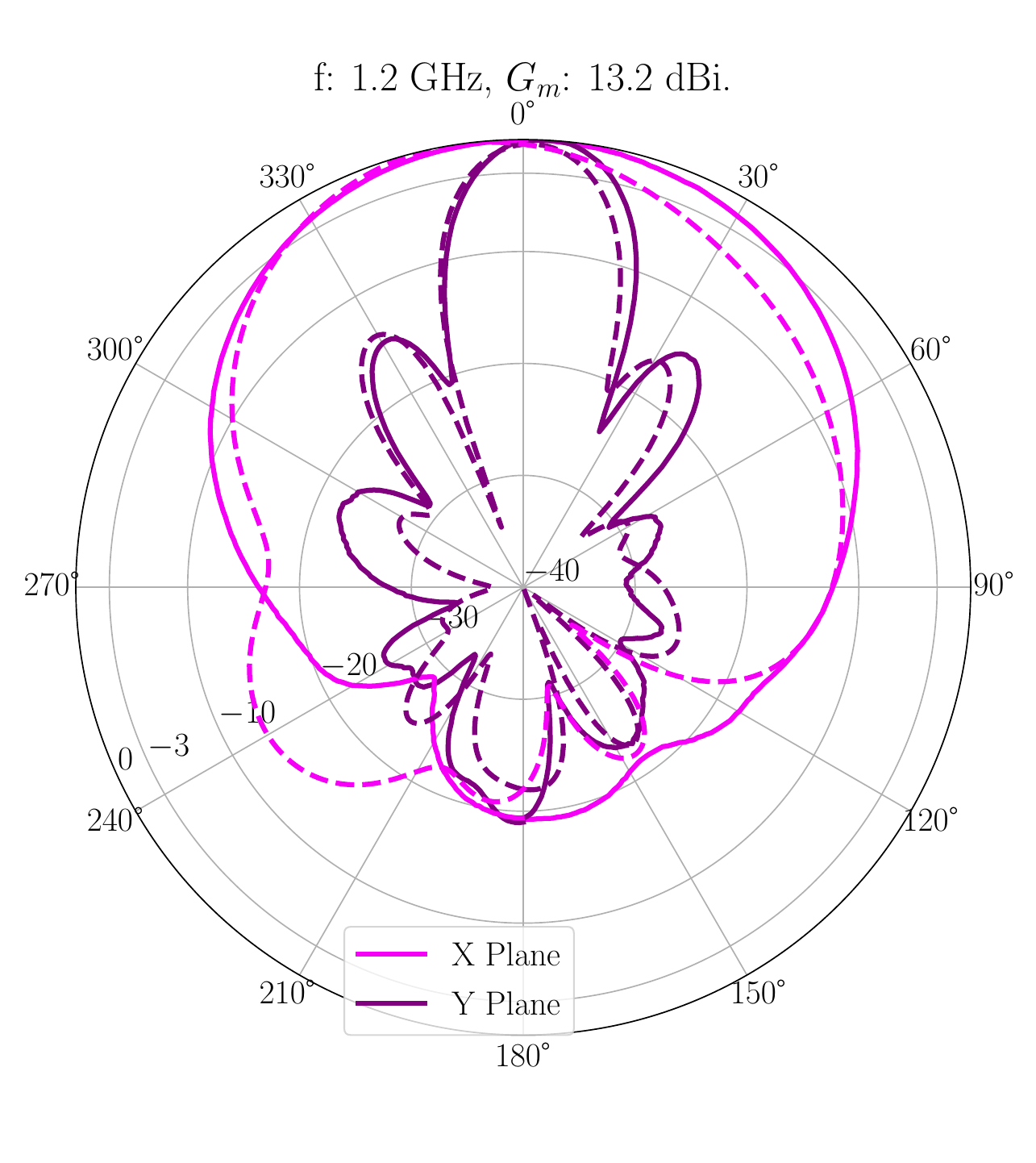}
         \caption{}
     \end{subfigure}
     \begin{subfigure}[b]{0.32\linewidth}
         \centering
         \includegraphics[width = 0.85\linewidth, trim = {14cm 0 13cm 0}, clip]{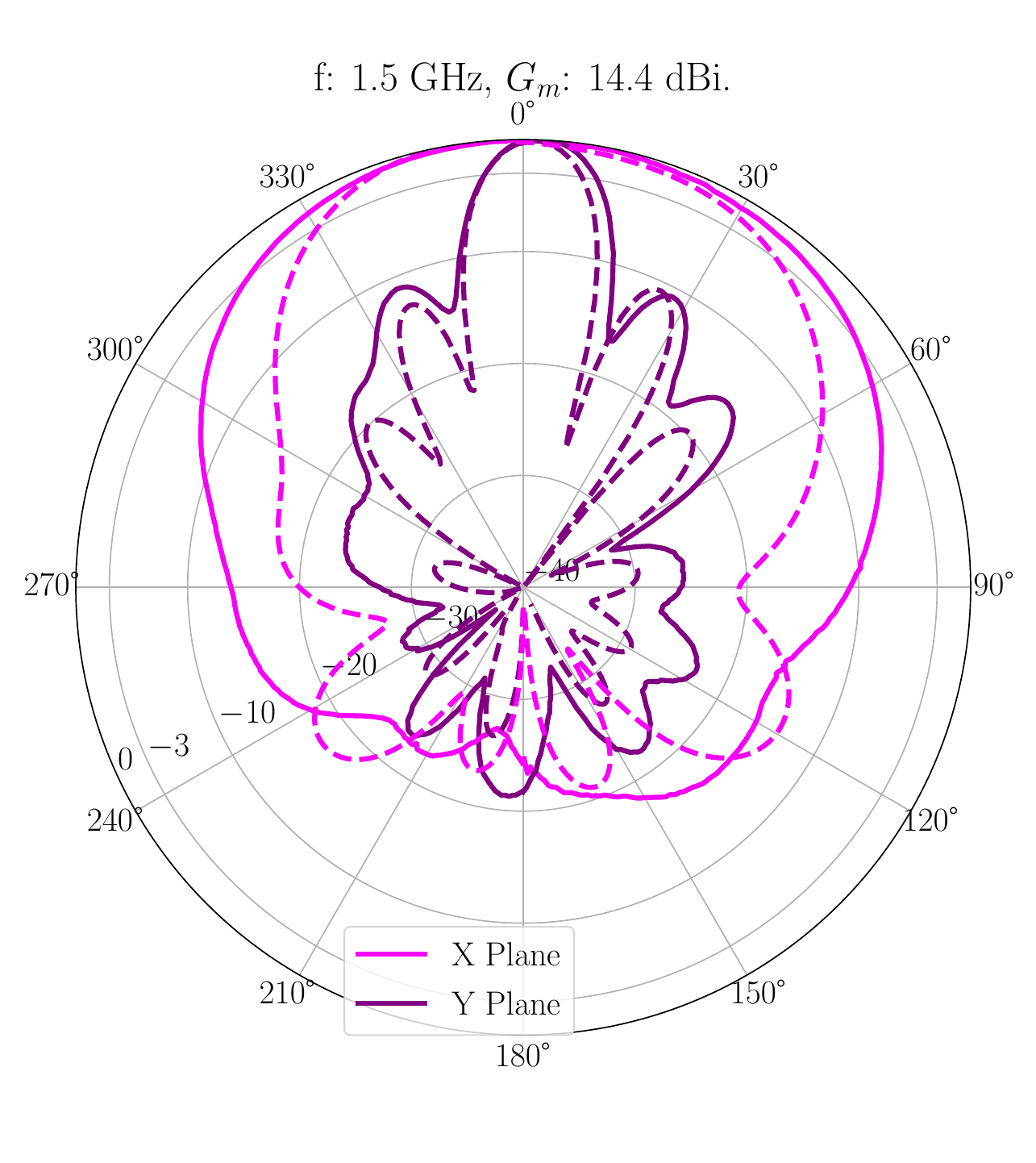}
         \caption{}
     \end{subfigure}
     \begin{subfigure}[b]{0.32\linewidth}
         \centering
         \includegraphics[width = 0.85\linewidth, trim = {14cm 0 13cm 0}, clip]{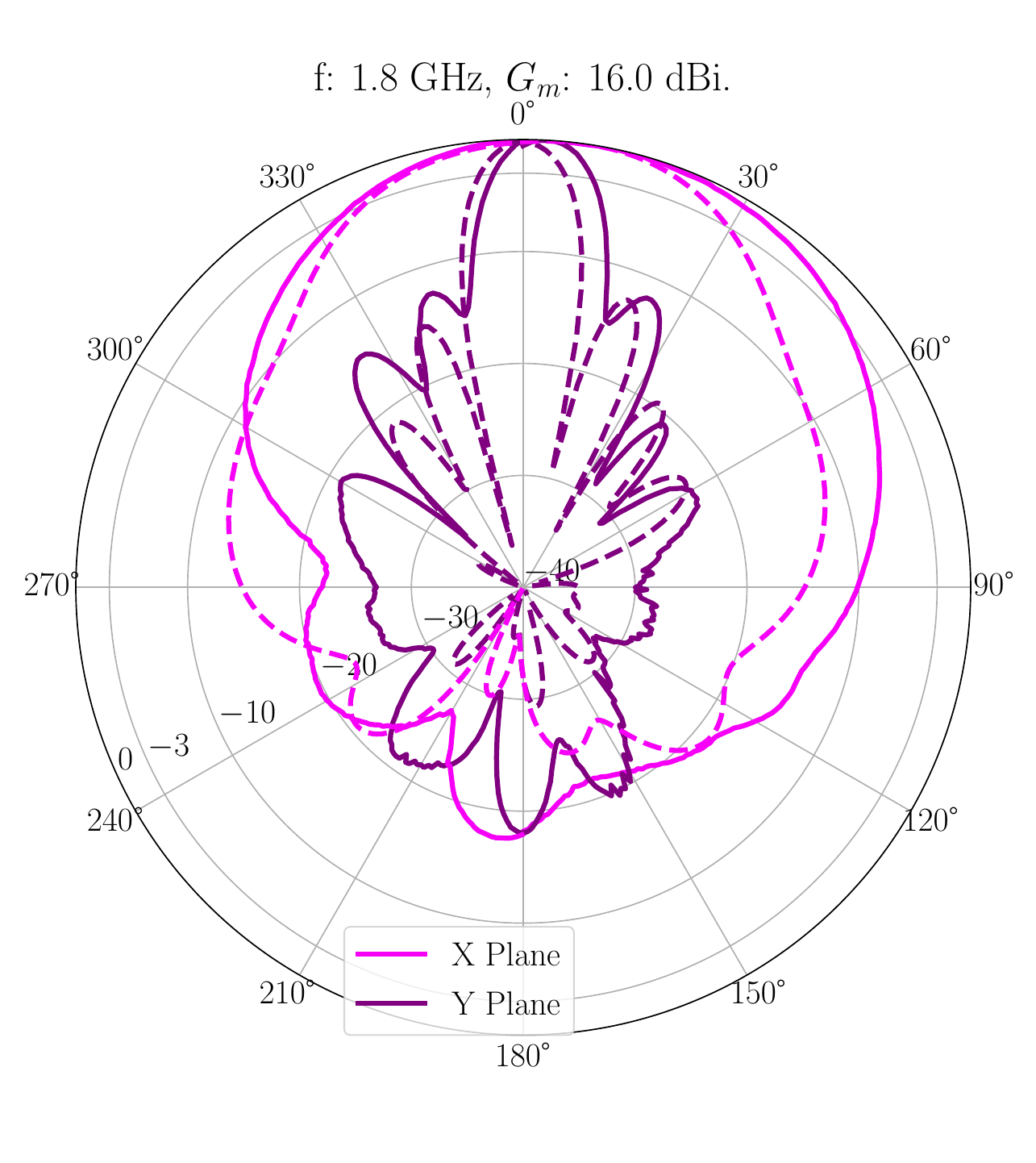}
         \caption{}
     \end{subfigure}
     \caption{Measured (solid lines) and simulated (dashed lines) radiation patterns of the L-shaped antenna array considering the feeding of 8 consecutive elements. Measurements are normalized with respect to their maximum, and simulations are normalized with respect to their maximum ($G_m$).}
     \label{fig: radiation patterns}
\end{figure}

The measured and simulated beam patterns in two orthogonal planes are shown in Fig.~\ref{fig: radiation patterns}. Only total gain measurements were made since the antenna is symmetric for both polarizations. We have selected 1200, 1500, and 1800~MHz as representative frequencies from the operation range. The simulated maximum gain and the frequency are indicated above each graph, while the measured and simulated HPBWs and Front-To-Back Ratios are shown in Table~\ref{table: meas pat}. It has to be noted that, since for detection we are going to use two consecutive sub-arrays in the $y$-axis (refer to Section~\ref{ss: doa}), the measurements were made by combining the eight successive elements of the $y$-axis.

Regarding the sidelobes, a good match is evident between the simulation and the actual measurements, both in amplitude and position. While the measured values of the sidelobes with the highest amplitude exceed the simulated values, they remain below 10~dB relative to the main lobe. This deviation is deemed acceptable for the proper functioning of the antenna. We can observe minor differences due to a small deviation in the pattern, which the mechanical tolerances of the holding structures can explain.

The measured front-to-back ratio is approximately 20~dB across the frequency range, well above the previously established 10~dB criterion. If we follow an analysis similar to that presented in Section~\ref{ss: fbr}, we can conclude that with this value of the front-to-back ratio, the ground pick is approximately 3~K, which is a negligible contribution to the typical values of receivers' noise temperatures.

Concerning the antenna aperture efficiency, the observed gain of 14.4~dBi at 1.5~GHz corresponds to an effective area of 0.088~m$^2$, while the physical area measures 0.141~m$^2$. This results in an aperture efficiency of 62\%. This low value is expected since the antenna consists mainly of dipoles. Despite its lower aperture efficiency, it is worth noting that the antenna offers the advantage of increased mobility with a standard equatorial mount, which is a significant benefit for the initial setup of ARTE.

Remembering that the array is planned to be positioned inside a dome, we must mention that the antenna's positioning ensures that the main lobe is wholly contained within the RF window, avoiding interaction with the opaque walls of the dome. However, the antenna's placement within the dome will inevitably affect the radiation pattern's shape. Measuring the radiation pattern of the system integrated into the dome is planned for future work. To accomplish this, we'd like to position a probe antenna at the zenith of the dome and rotate the array using the equatorial mount to measure the pattern.

\begin{figure}[t!]
    \centering
    \includegraphics[width = 0.5\linewidth]{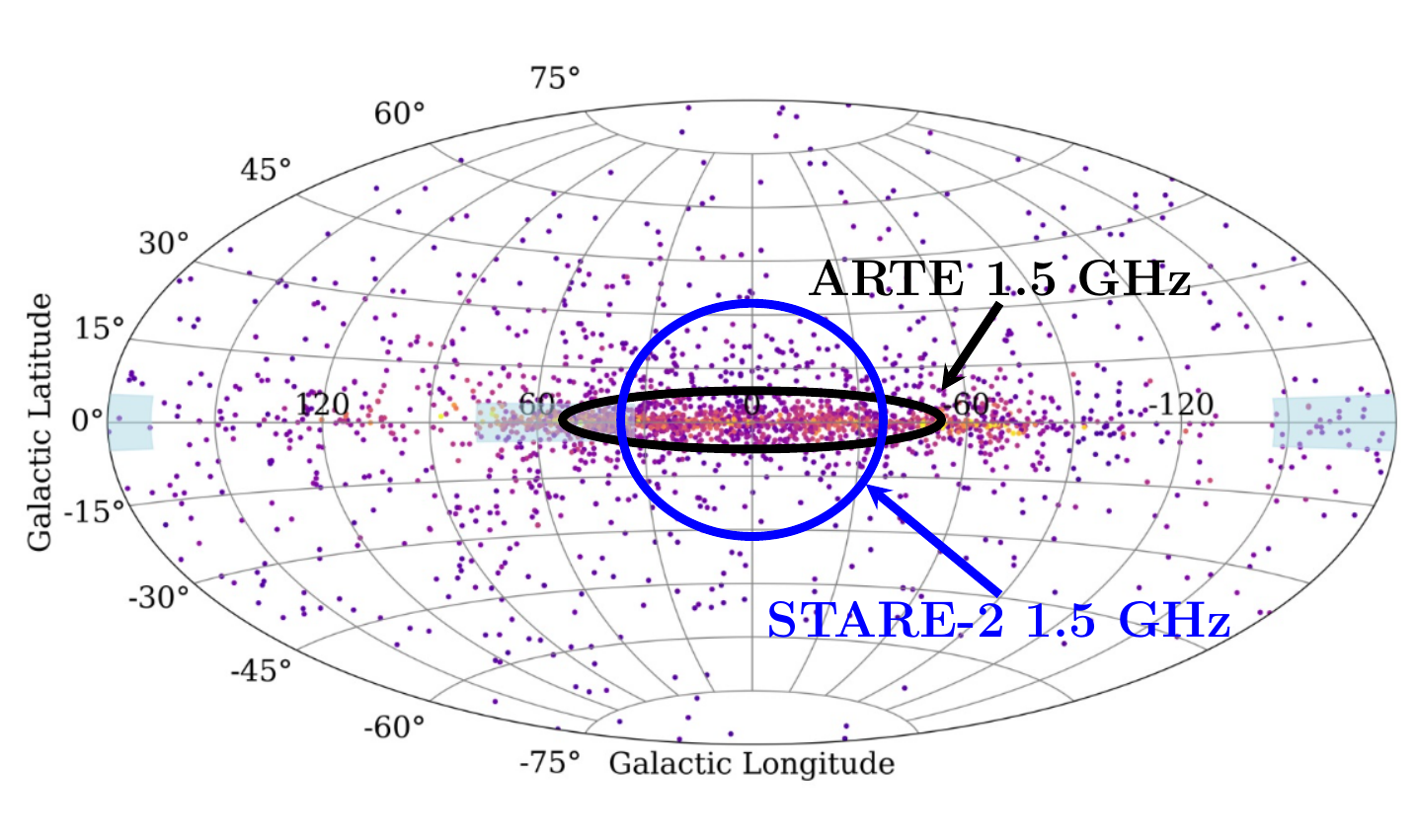}
    \caption{Comparison between STARE-2 and the ARTE antenna array at 1.5~GHz. Similarly to Fig.~\ref{fig: pat shape}, the points represent the distribution of pulsars, the ellipses correspond to the projection of the radiation patterns, and the axes of the ellipses are the HPBWs.}
    \label{fig: stare2 comparison}
\end{figure}

Finally, as ARTE aims to be a more sensitive version of STARE-2, it is essential to note the difference in observing the Galaxy with both antennas. The projections of the radiation pattern of the STARE-2 and ARTE antennas are shown in Fig.~\ref{fig: stare2 comparison}. As can be seen, the ARTE beam is more concentrated in the area of interest, implying a sensitivity increase of approximately 3.35 at 1.5~GHz due only to the beam's shape. Section~\ref{ss: sensitivity} shows consideration of factors other than the beam shape, such as the ground pick of the back lobe, losses in cables and combiners, and the addition of the receiver noise temperature.

\begin{table}[t!]
    \caption{Simulated and Measured radiation characteristics of the antenna.}
    \centering
    \scalebox{0.95}[0.95]{%
    \begin{tabular}{l|c|c|c|c|c|c}
    \hline \hline
    & \multicolumn{2}{c|}{\textbf{1200~MHz}} & \multicolumn{2}{c|}{\textbf{1500~MHz}} & \multicolumn{2}{c}{\textbf{1800~MHz}} \\ \hline
    \textbf{Parameter} & Sim. & Meas. & Sim. & Meas. & Sim. & Meas. \\ \hline
    HPBW$_{x}$ & 54$^\circ$ & 80$^\circ$ & 70$^\circ$ & 97.5$^\circ$ & 55$^\circ$ & 80$^\circ$ \\
    HPBW$_{y}$ & 17$^\circ$ & 20$^\circ$ & 13$^\circ$ & 15$^\circ$ & 12.5$^\circ$ & 15$^\circ$ \\ 
    Front-to-Back Ratio & 22~dB & 19~dB & 26~dB & 21~dB & 29~dB & 18~dB \\ \hline \hline 
    \end{tabular}
    }
    \label{table: meas pat}
\end{table} 

% \textcolor{red}{STARE2 is sensitive to one-millisecond transients above $\sim$ 300 kJy while ARTE is sensitive to a one-millisecond transients above  $\sim$ 67.6 kJy.}
%%%%%%%%%%%%%%%%%%%%%%%%%%%%%%%%%%
%%%%%%%%%%%%%%%%%%%%%%%%%%%%%%%%%%
\subsection{Sensitivity and Fluence Calculations}
\label{ss: sensitivity}
%%%%%%%%%%%%%%%%%%%%%%%%%%%%%%%%%%
%%%%%%%%%%%%%%%%%%%%%%%%%%%%%%%%%%

\begin{figure}[t!]
    \centering
    \includegraphics[width = 1\linewidth]{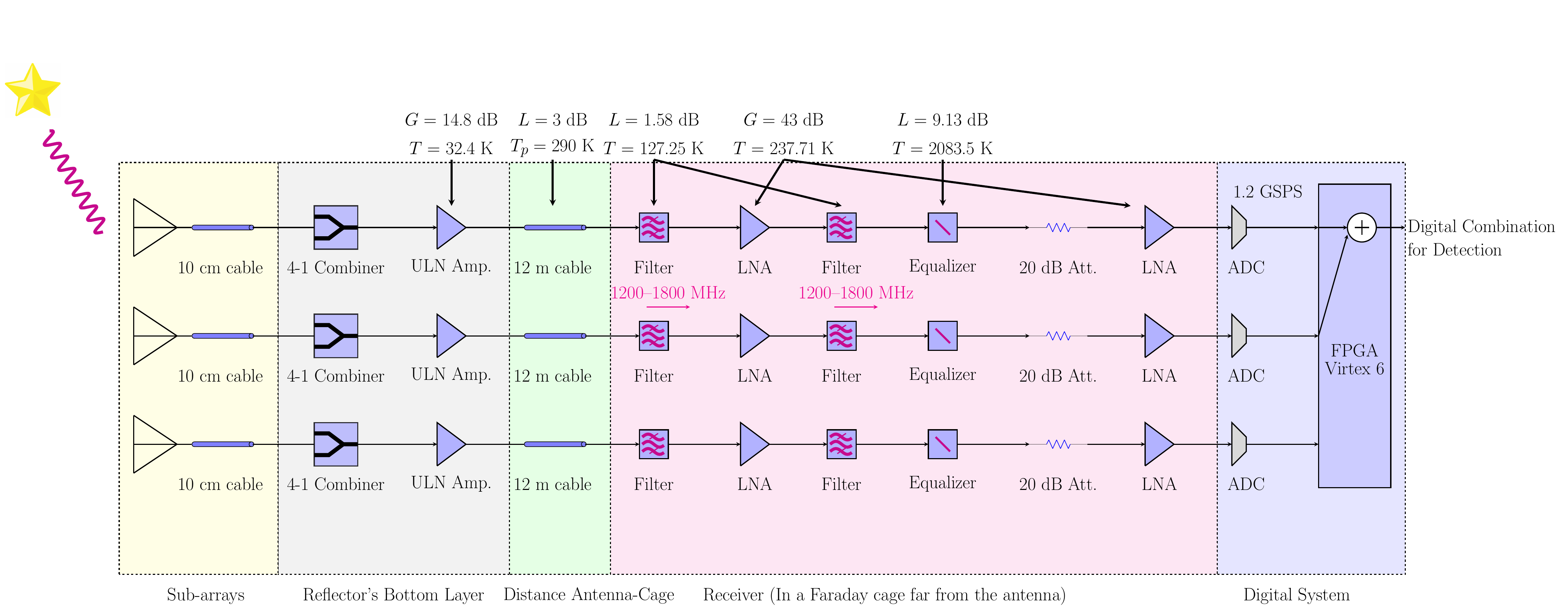}
    \caption{Proposed receiver for integrating the proposed antenna array in ARTE. The choice of commercial components is beyond the scope of this article, but it is mandatory to estimate the minimum fluence that can be achieved with the proposed antenna.}
    \label{fig: system}
\end{figure}

To make a reliable comparison with STARE-2, we shall estimate the minimum fluence our antenna can detect when integrated with a receiver. We will consider the receiver proposed in Fig.~\ref{fig: system} to do so, which considers the use of real commercial components. This receiver is composed as follows: To minimize losses, we will consider integrating the power combiners and ultra-low noise (ULN) amplifiers immediately adjacent to the bottom layer of the reflecting plane. This implies the utilization of 10~cm cables between the antennas and the combiners. Then, to connect the antenna at the top of the dome with the rest of the receiver located far away in a Faraday cage, 12-meter-long coaxial cables with 3~dB of total loss will be used. The two 1200--1800~MHz bandpass filters shown in the figure help to obtain greater rejection outside the band of interest, while the 20~dB attenuator that precedes the last amplifier is used not to saturate the latter.

To calculate the minimum fluence F of our system, expressed in kJy$\cdot$ms, we use the radiometer equation \cite{lorimer radio},

\begin{equation}
\mathrm{F = \Gamma ^{-1}\frac{SNR \cdot T_{sys} \cdot \tau}{\sqrt{\Delta \upsilon \cdot \tau }}}, \label{ec: fluence}
\end{equation}
where $\Gamma$ is the sensitivity (in K Jy$^{-1}$), $\Delta \upsilon$ is the bandwidth of interest, that in our case is 600~MHz, $\tau$ is the time resolution (10~ms), SNR is the minimum signal-to-noise ratio (7~dB, similar to the one considered in STARE-2), and $T_{\textup{sys}}$ is the noise temperature of the system, that comes from the sum of the receiver noise temperature $T_{\textup{rx}}$ and the antenna temperature $T_{\textup{A}}$.

Assuming losses of 0.5~dB in the combiners and 0.5~dB in the cables, we have an equivalent 1~dB reduction in antenna gain. Then, the sensitivity $\Gamma$ for the ARTE antenna, expressed in K Jy$^{-1}$, is calculated as

\begin{align}
\mathrm{\Gamma = \frac{A_e}{2k_B}},
\label{sensitivity}
\end{align}
where $\mathrm{A_e}$ is the effective area and $\mathrm{k_B}$ is the Boltzmann constant~\cite{sensitivity}. After updating the gain values in Fig.~\ref{fig: radiation patterns} with the 1~dB losses of cables and combiners, we obtain the sensitivity values shown in Table~\ref{table: sensitivity_fluence}. To calculate $T_{\textup{sys}}$, we must consider that the antenna is connected to the receiver through a lossy transmission line (combiners and 10~cm cables). Therefore, we shall consider the antenna temperature as~\cite{pozar}

\begin{table}[t!]
    \caption{Values of sensitivity $\Gamma$ and fluence F.}
    \centering
    \scalebox{0.95}[0.95]{%
    \begin{tabular}{c|c|c|c}
    \hline \hline
    \textbf{Parameter}  & \textbf{1200~MHz} & \textbf{1500~MHz } & \textbf{1800~MHz} \\ \hline
    $\Gamma $  (K Jy$^{-1}$)   & 3 $\cdot 10^{-5}$ & 2.5 $\cdot 10^{-5}$ & 2.5 $\cdot 10^{-5}$  \\ 
    Fluence with FBR $=$ 20~dB (kJy~$\cdot$~ms)   &  109.4 &  131.3 &  131.3 \\
    Fluence with FBR $=$ 10~dB (kJy~$\cdot$~ms)   &  120.3 &  144.4 &  144.4 \\ \hline \hline 
    \end{tabular}
    }
    \label{table: sensitivity_fluence}
\end{table} 

\begin{align}
T_{\textup{A}} = \frac{1 - |S_{11}|^2}{L} T_b + \frac{L - 1}{L} \left(1 + \frac{|S_{11}|^2}{L} \right)T_p,
\label{sensitivity}
\end{align}
where $L = 1$~dB is the loss factor of the lossy line, $T_b$ is the brightness temperature seen by the antenna, composed of 30~K from the Galactic center seen by the main lobe, and 3~K pick up from the back lobe (considering the measured front-to-back ratio of $\approx$~20~dB), $T_p = 290$~K is the physical temperature, and $|S_{11}|^2 = 10$~dB is the worst reflection coefficient of the antenna. We obtain $T_{\textup{A}} = 88$~K using these values. Finally, to calculate the value of $T_{\textup{rx}}$, we used the values of gain, loss, and noise temperature of Fig.~\ref{fig: system} and the cascaded Friis equation to obtain $T_{\textup{rx}} = 73$~K, and therefore $T_{\textup{sys}} = 161~$K. Replacing all these values in~\eqref{ec: fluence}, we get the minimum fluences shown in the middle row of Table~\ref{table: sensitivity_fluence}. Considering that STARE-2 is sensitive to one-millisecond transients above $\sim$ 300~kJy while ARTE is sensitive to one-millisecond transients above $\sim$ 131.3~kJy at 1.5~GHz, we conclude a sensitivity increase of 2.28. Additionally, we can consider a ground pick of 26~K -equivalent to the FBR of 10~dB- for greater completeness in calculating the minimum fluence. We do this because, although the measured FBR was 20~dB, incorporating the antenna in the dome can alter this value. Considering a ground pick of 26~K, we obtain $T_{\textup{A}} =  104$~K, $T_{\textup{sys}} = 177~$K, and the fluence values shown in the last row of Table~\ref{table: sensitivity_fluence}.

%% file: conclus.tex
\section{Conclusions}
\label{s: conclus}

This paper presents a broadband, compact, half-space antenna array designed for detecting and localizing FRBs within the Milky Way. It consists of 3 sub-arrays placed in an L-shape for source localization. Each sub-array consists of 4 dual-pol antenna elements, which allows the radiation pattern to be shaped to match the form of the Milky Way. Moreover, the miniaturized size of the antenna elements provides array operation without phase ambiguities. 

The unique antenna attributes outlined in this research make it an ideal choice for developing a cost-effective telescope to survey the Milky Way for FRBs. We estimate an increment in sensitivity of 2.28 when compared to using a horn antenna as the one used in STARE-2 to cover a similar area of the Milky Way.

\section*{Acknowledgements}

We gratefully acknowledge support of ANID funds Basal FB210003, FONDEF ID21-10359 and Fondecyt 1221662. We also thank Universidad Tecnica Federico Santa Maria for access to its anechoic chamber.